%% file: ms.tex
\documentclass[a4paper,cleveref,nosubfigcap,english]{lipics-v2021}
\hideLIPIcs
\nolinenumbers

\usepackage[T1]{fontenc}

\usepackage{amsmath,amssymb}
\usepackage{hyperref}
\usepackage{url}
\usepackage{booktabs}
\usepackage{siunitx}
\usepackage{subcaption}
\usepackage{tikz}
\usepackage{xcolor}
\usepackage{placeins}

\crefname{equation}{}{}

\title{Banana Trees for the Persistence in Time Series Experimentally}
\titlerunning{Banana Trees for the Persistence in Time Series Experimentally}

\author{Lara Ost}
{Faculty of Computer Science, Doctoral School Computer Science, University of Vienna, Austria}
{lara.ost@univie.ac.at}
{https://orcid.org/0000-0003-4311-9928}
{Supported by the Vienna Graduate School on Computational Optimization (VGSCO), FWF project no.~W1260-N35.}
\author{Sebastiano {Cultrera di Montesano}}
{Eric and Wendy Schmidt Center, Broad Institute of MIT and Harvard, Cambridge, USA}
{scultrer@broadinstitute.org}
{https://orcid.org/0000-0001-6249-0832}
{Supported by the Eric and Wendy Schmidt Center at the Broad Institute of MIT and Harvard.}
{}
\author{Herbert Edelsbrunner}
{IST Austria (Institute of Science and Technology Austria), Kloster\-neu\-burg, Austria}
{herbert.edelsbrunner@ist.ac.at}
{https://orcid.org/0000-0002-9823-6833}
{Partially supported by the Wittgenstein Prize, FWF grant no.~Z~342-N31, and by the DFG Collaborative Research Center TRR~109, FWF grant no.~I~02979-N35.}
\funding{}

\authorrunning{L. Ost, S. Cultrera di Montesano, H. Edelsbrunner}
\Copyright{Lara Ost, Sebastiano Cultrera di Montesano and Herbert Edelsbrunner}


\supplement{The source code is available as follows:}
\supplementdetails[swhid={swh:1:dir:8407d6500f6dbccdb5102f3de3db35a9160b79c8}]{Software}{www.github.com/laraost/BananaPersist}

\EventEditors{Oswin Aichholzer and Haitao Wang}
\EventNoEds{2}
\EventLongTitle{41st International Symposium on Computational Geometry (SoCG 2025)}
\EventShortTitle{SoCG 2025}
\EventAcronym{SoCG}
\EventYear{2025}
\EventDate{June 23--27, 2025}
\EventLocation{Kanazawa, Japan}
\EventLogo{socg-logo.pdf}
\SeriesVolume{332}
\ArticleNo{XX}     %

\newcommand{\refinsubcaption}[1]{\textbf{\subref{#1}}}

\newcommand{\sidecaption}[1]%
{\raisebox{\abovecaptionskip}{\begin{subfigure}[t]{0.5em}
  \caption[singlelinecheck=off]{}%
  \label{#1}
\end{subfigure}}\ignorespaces}

\newcommand{\flfrac}[2]{\left\lfloor\frac{#1}{#2}\right\rfloor}

\newcommand{\Nspace}{\mathbb{N}}
\newcommand{\Zspace}{\mathbb{Z}}
\newcommand{\Rspace}{\mathbb{R}}
\newcommand{\eps}{\varepsilon}

\newcommand{\normaldist}{\mathcal{N}}

\newcommand{\banana}[2]{B(#1,#2)}
\newcommand{\inptr}{\texttt{in}}
\newcommand{\midptr}{\texttt{mid}}
\newcommand{\upptr}{\texttt{up}}
\newcommand{\dnptr}{\texttt{dn}}
\newcommand{\lowptr}{\texttt{low}}
\newcommand{\dthptr}{\texttt{dth}}

\newcommand{\Gudhiname}{\texttt{Gudhi}}

\newcommand{\rwfunc}{r}

\newcommand{\wclmfunc}{w_\ell}
\newcommand{\wctmfunc}{w_t}

\newcommand{\cpp}[1]{C\texttt{++}#1}

\usetikzlibrary{math}

\definecolor{oiorange}{RGB}{230,159,0}
\definecolor{oicyan}{RGB}{86,180,233}
\definecolor{oigreen}{RGB}{0,158,115}
\definecolor{oiyellow}{RGB}{240,228,66}
\definecolor{oiblue}{RGB}{0,114,178}
\definecolor{oired}{RGB}{213,94,0}
\definecolor{oipurple}{RGB}{204,121,67}

\newcommand{\drawemptybanana}[2]{
    \draw[bend right] (#1) to (#2);
    \draw[bend left] (#1) to (#2);
}

\usepackage[colorinlistoftodos,prependcaption,textsize=tiny]{todonotes}

\begin{document}

\maketitle

\begin{abstract}
  In numerous fields, dynamic time series data require continuous updates, necessitating efficient data processing techniques for accurate analysis. This paper examines the banana tree data structure, specifically designed to efficiently maintain the multi-scale topological descriptor commonly known as \emph{persistent homology}
  for dynamically changing time series data. We implement this data structure and conduct an experimental study to assess its properties and runtime for update operations. Our findings indicate that banana trees are highly effective with unbiased random data, outperforming state-of-the-art static algorithms in these scenarios. Additionally, our results show that real-world time series share structural properties with unbiased random walks, suggesting potential practical utility for our implementation.
\end{abstract}

\section{Introduction}
\label{sec:1}

Time series are pervasive across numerous disciplines, ranging from economics and finance to environmental science and healthcare. 
Often, time series data is dynamic, not static; for example, wearable devices continuously monitor health metrics such as heart rate or blood sugar levels, requiring robust techniques for instant analysis and decision-making. 
Effective synthesis of the underlying trends in such dynamic data is crucial for predictive modeling and strategic interventions.

\smallskip
Within the field of topological data analysis, persistent homology \cite{Car09,EdLeZo02} is recognized as a powerful framework for capturing multi-scale features in complex datasets. 
Researchers have long considered the challenge of dynamic persistence, and the vineyard algorithm \cite{CEM06} was the first developed to address updates in persistence diagrams (defined in Section \ref{sec:2.1}) as the input data evolves; see \cite{LuNe21} for an implementation.
This algorithm, while capable of handling data more general than just time series, requires time linear in the size of the input complex per update, which is restricted to swapping the order of two values.
This prompts us to question whether faster processing could be achieved for one-dimensional data.
To address this challenge, we implement the banana tree data structure \cite{CEHO24}, recently introduced and tailored specifically for the persistent homology of dynamic time series. 
Their theoretical analysis promises the processing of each update in time logarithmic in the number of items plus linear in the number of changes in the persistent diagram; compare this with the linear time require to recompute the diagram \cite{Gli23}. 
This paper investigates what these results mean in practice.
We highlight some of the specific contributions:
\medskip \begin{itemize}
    \item \emph{Efficiency at scale:} 
    we demonstrate that for large datasets, such as those containing over $10^6$ items, performing a local or topological update on an unbiased random walk using banana trees is at least 100 times faster than recomputing persistence with \Gudhiname{} \cite{Gudhi}, the state of the art static algorithm (see Figure \ref{fig:maintenance-speedup}).
    \item \emph{Structure and performance:} we explore how the structure of banana trees, characterized by parameters such as the number of critical items, the nesting depth of bananas, and the lengths of trails, directly impacts the running time of our algorithms. This analysis identifies the types of data best suited for efficient processing (see Figure \ref{fig:structure}).
    \item \emph{Worst-case scenarios:} we construct specific examples that challenge our algorithms, and show empirically that these scenarios are rare and highly unstable, reinforcing the robustness of our approach (see Table~\ref{tbl:topological-wc-speedup}).
    \item \emph{Real-world data:} through analysis of three specific datasets, we illustrate that real-world data sometimes exhibits structural properties similar to those of unbiased random walks. 
    Preliminary evidence suggests potential for the broad applicability of banana trees in practical scenarios (see Table \ref{tbl:real-world-data}).
\end{itemize} \medskip
In software for topological data analysis, tools like the \Gudhiname{} library~\cite{Gudhi}, Dionysus~\cite{Mor23.1,Mor23.2}, and Ripser~\cite{Bauer21} are established for static datasets but require complete recomputation for updates, making them less suited in dynamic settings. 
We aim for our implementation to set a new standard in the topological processing of dynamic time series. 

\subparagraph*{Outline.} 
Section \ref{sec:2} introduces persistent homology tailored to time series data and gives an overview of the banana trees data structure.
Section \ref{sec:3} evaluates the performance of banana trees through experiments involving time series generated from random walks, with and without a bias. 
Section \ref{sec:4} explores three distinct types of input time series to assess the performance of banana trees: worst-case scenarios, quasi-periodic signals, and real-world data. 
Section \ref{sec:5} concludes the paper with a summary of our findings.

\section{Banana Trees for Time Series}
\label{sec:2}

We start by introducing persistent homology, a method to discern features of data across multiple scales \cite{EdHa10}. 
Forfeiting the generality of this theory, we focus on time series data. 
We also explain what banana trees are and how they relate to persistent homology; see \cite{CEHO24} for details on this data structure and its algorithms.

\subsection{Persistent Homology of Time Series}
\label{sec:2.1}

By a \emph{time series} we mean a linear list of real numbers, $c_0, c_2, \dots, c_{n-1}$, which we view as a piecewise linear map, $f \colon [0,n-1] \to \Rspace$, with $f(i) = c_i$ for $0 \leq i \leq n-1$.
We refer to $i$ as an \emph{item} and $c_i$ its \emph{value}.
A \emph{critical item} is a local minimum or a local maximum of this map, and all other items are \emph{non-critical}, e.g. item $i$ if $f(i-1) < f(i) < f(i+1)$.
To simplify the discussion, we assume that the map is \emph{generic}, by which we mean that its items have distinct values.
In this case, the endpoints (items $0$ and $n-1$) are necessarily critical.

\smallskip
The \emph{sublevel set} of $f$ at $t \in \Rspace$, denoted $f_t = f^{-1} (-\infty, t]$, are the points $x \in [0,n-1]$ that satisfy $f(x) \leq t$.
When $t$ passes the value of a minimum from below, then the number of connected components of $f_t$ increases by $1$, and if $t$ passes the value of a maximum from below, the number of connected components decreases by $1$, unless the maximum is an endpoint, in which case the number does not change.
Symmetrically, we call $f^t = f^{-1} [t,\infty)$ the \emph{superlevel set} of $f$ at $t$.
Observe that $f^t$ is the sublevel set of $-f$ at $-t$, and that the minima and maxima of $-f$ are the maxima and minima of $f$.
Persistent homology tracks the evolution of the connected components while the sublevel set of $f$ grows, and formally defines when a component is born and when it dies.
Complementing this with the same information for the superlevel sets of $f$, we get what is formally referred to as \emph{extended persistent homology}; see \cite{CEH09} for details.
It is best constructed in two phases:
\medskip \begin{itemize}
  \item In \emph{Phase One}, we track the connected components of the sublevel set, $f_t$, as $t$ increases from $-\infty$ to $\infty$.
  A component is \emph{born} at the smallest value of $t$ at which a point of the component belongs to $f_t$, which is necessarily a minimum. 
  The component \emph{dies} when it merges with another component that was born earlier, which is necessarily at a maximum in the interior of $[0,n-1]$.
  The \emph{ordinary subdiagram} of $f$ records the birth and death of every component with a point in the plane whose abscissa and ordinate are those values of $t$ at which the component is born and dies, respectively; see Figure~\ref{fig:windows-and-bananas}.
  \item In \emph{Phase Two}, we track the connected components of the superlevel set, $f^t$, as $t$ decreases from $\infty$ to $- \infty$.
  Birth and death are defined accordingly, and the components are recorded in the \emph{relative subdiagram}.
\end{itemize} \medskip
By construction, the points in the ordinary subdiagram lie above and those of the relative subdiagram lie below the diagonal.
The component born at the global minimum of $f$ is special because it does not die during Phase~One.
Instead, it dies at the global minimum of $-f$, which is the global maximum of $f$.
In topological terms, this happens because the one connected component still alive at the beginning of Phase~Two dies in relative homology when its first point enters the superlevel set.
This class is represented by the sole point in the \emph{essential subdiagram}.
The \emph{extended persistence diagram} is the disjoint union of the three subdiagrams.
Hence, the diagram is a multi-set of points in $\Rspace^2$, and so are the three subdiagrams, unless the map is generic, in which case the diagram is a set.
See the left panel in Figure~\ref{fig:windows-and-bananas} for an example but note that it only shows a small number of points in the ordinary subdiagram.
An important property of persistence diagrams is their stability with respect to small perturbation of the input data, which was first proved in \cite{CEH07}.

\smallskip
The points in the persistence diagram can be characterized using the concept of a \emph{window} introduced in \cite{BCES21}:
start with a rectangular frame spanned by a minimum and a maximum in the graph of $f$ such that the minimum lies on the lower edge, and extend the frame horizontally as long as the graph does not intersect the upper edge in its interior.
Call the initial frame the \emph{mid-panel}, its extension the \emph{in-panel}, and the symmetric extension for $-f$ the \emph{out-panel}.
As proved in \cite{BCES21}, the min-max pair defines a point in the persistence diagram iff the graph of $f$ reaches the lower edge in the out-panel.
In this case, we call the twice extended frame a \emph{triple-panel window}.
The corresponding \emph{double-panel window} consists of the mid-panel and the in-panel but not the out-panel.
Any two double-panel windows of $f$ are either disjoint or nested, a property not shared by the triple-panel windows.
We use this property to \emph{augment} the persistence diagram by drawing an arrow from a point to another, if the double-panel windows of the first point is nested inside the double-panel window of the other point, without any other window being nested between them.
See Figure~\ref{fig:windows-and-bananas}, which shows four double-panel windows and the corresponding points and arrows in the augmented persistence diagram.
The displayed frames are indeed windows because each has an out-panel---next to the mid-panel and thus opposite to the in-panel---which extends as far as the horizontal projection of the minimum (a maximum of $-f$) to the graph of the function.

\subsection{Banana Trees}
\label{sec:2.2}

We sketch the banana trees while referring to \cite{CEHO24} for the details.
Suffice to say that they are based on the Cartesian tree introduced by Vuillemin~\cite{Vui80}, which stores a list of items in order such that each path from a leaf to the root is ordered by value.
This tree has a unique decomposition into paths that correspond to double-panel windows, and the banana tree splits each path into two trails representing the windows nested inside the in-panel and the mid-panel of the corresponding window.
\begin{figure}[htb]
    \centering
    \vspace{0.1in}
    \includegraphics[height=1.5in]{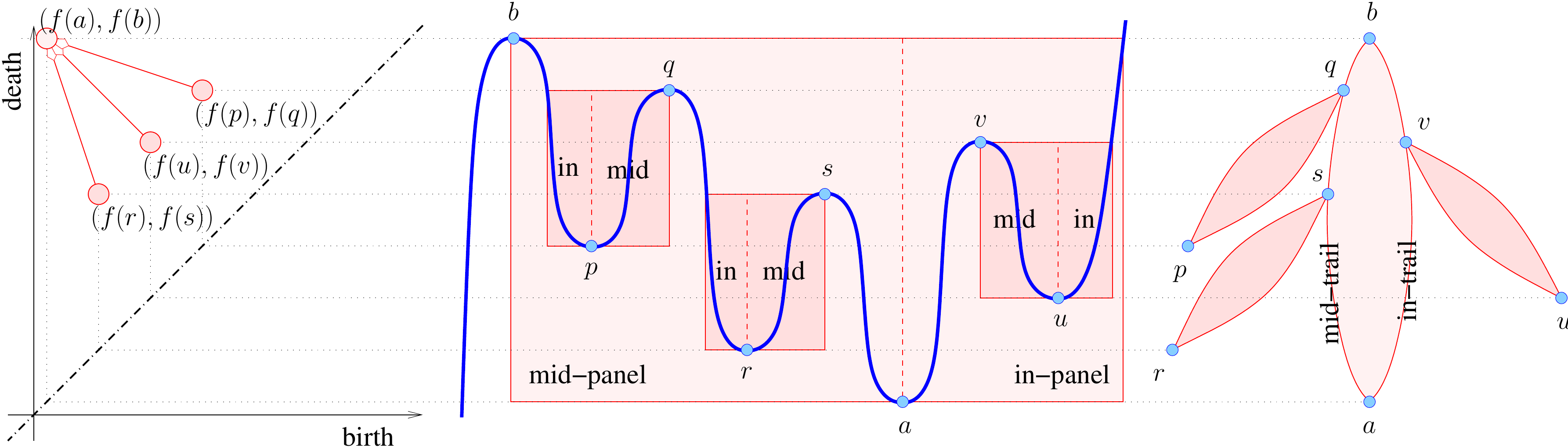}
    \caption{\footnotesize
      \emph{Middle:} a piece of the graph of $f$ with four double-panel windows of which three are nested inside the fourth.
      \emph{Left:} the corresponding points in the persistence diagram and the arrows that reflect the nesting relation among the windows.
      \emph{Right:} the four corresponding bananas in the up-tree of $f$.}
    \label{fig:windows-and-bananas}
\end{figure}

A time series represented by a piecewise linear function, $f$, is stored in two banana trees together with a linked list and two standard dictionaries.
We call the first banana tree the \emph{up-tree} of $f$ because it corresponds to Phase One of the persistence diagram construction.
The second banana tree is the \emph{down-tree} of $f$, which is the up-tree of $-f$.
Because of this symmetry, it suffices to describe the up-tree of $f$.
Its \emph{leaves} are the minima of $f$, and its \emph{internal nodes} are the maxima of $f$, with the exception of endpoint maxima, if they exist, since they are not critical in Phase One.
Each point in the ordinary persistence diagram is a min-max pair, $(a,b)$, and it is represented by a \emph{banana} that connects the leaf $a$ to the internal node $b$ with two parallel trails.
The \emph{mid-trail} is a doubly-linked list connecting $a$ to $b$ with the maxima that span windows nested in the mid-panel of the window of $a,b$, and the \emph{in-trail} is symmetrically defined; see the right panel in Figure~\ref{fig:windows-and-bananas}.
The tree structure arises because the maxima belong to the similarly defined bananas of the nested windows.
Each trail is sorted in the order of value but also in the order of location along the time series.
For example, the mid-trail of the big banana in Figure~\ref{fig:windows-and-bananas} stores $a, s, q, b$, which monotonically increases in value and monotonically decreases in location.
Compare this with the in-trail storing $a,v,b$, which monotonically increases in value and in location after we discard $b$.
Indeed, we think of $b$ as part of the mid-trail while only connecting the in-trail to the rest of the tree without properly belonging to it.
The global minimum of $f$ (not shown in Figure~\ref{fig:windows-and-bananas}) is special because it is not paired in Phase One.
Indeed, it is the extra leaf whose path upward does not end at an internal node.
We therefore add a \emph{special root} as the parent of the root, connected to the global minimum by the \emph{root banana}.
While there is only one root banana, there are possibly many \emph{leaf bananas}, which are the bananas with two empty trails (beside the minimum and maximum they connect).
Indeed, we call the number of nodes strictly between the minimum and maximum the \emph{length} of the trail, which for trails of leaf bananas is necessarily zero.
Another important concept is the \emph{nesting depth} of a banana, which is the number of windows that contain the corresponding window.
The root banana has nesting depth zero, and all other bananas have positive nesting depth.

\smallskip
Each update to a time series stored in a banana tree reduces to a sequence of elementary operations, which we name by their impact on the function, $f$.
An \emph{interchange} happens when two maxima or two minima swap the order of their values.
Unless their locations are near each other, there is a good chance that an interchange does not have any effect on the banana tree.
Otherwise, it is akin a rotation in a binary search tree, if the interchange is between two maxima, and akin a local rearrangement of the path decomposition, if the interchange is between two minima.
A \emph{cancellation} removes a min-max pair or, equivalently, the corresponding leaf banana,
an \emph{anti-cancellation} is the inverse of a cancellation,
and a \emph{slide} swaps the type of a critical item with that of a neighboring non-critical item.
The latter three operations appear only a constant number of times whenever we adjust a value.
Nevertheless, anti-cancellations present challenges to fast implementation as we have to find out where the new banana is to be inserted,
and this search is not supported by any special purpose data structure.
On the other hand, multiple interchanges may happen in sequence, so it is important to charge the time for processing each to a change in the persistence diagram.
We refer to \cite{CEHO24} and \cref{app:A} for details.

\section{Experimental Results for Random Walks}
\label{sec:3}

This section studies the structural properties of banana trees for random walks, both biased and unbiased.
In addition, it compares the running times of local and topological operations with the static algorithm in \Gudhiname{} \cite{Gudhi}.
We do not compare it with vineyards \cite{CEM06}, which lack optimized software for the one-dimensional case.

\smallskip
We implemented the banana trees data structure in \cpp{20}, using the AVL trees and splay trees provided by \texttt{boost.intrusive} \cite{Boost}.
The former offer good performance in general, the latter allow for simple splitting and joining.
Hence, we use splay trees when performing topological operations and AVL trees everywhere else.
The experiments were run on a machine with two AMD EPYC 9534 64 core CPUs and 1.5 terabyte of main memory.

\subsection{The Experimental Set-up}
\label{sec:3.1}

We write $\gamma \sim \normaldist(\mu, \sigma)$ for a normally distributed random variable with mean $\mu$ and standard deviation $\sigma$.
Given $\mu, \sigma$, a \emph{random walk} of length $n$ is a function, $\rwfunc = \rwfunc_{\mu, \sigma}$, from the first $n$ non-negative integers to the reals, defined by
\begin{align}
  \rwfunc (0) = 0 \mbox{\rm ~~and~~} r(i) = r(i-1) + \gamma_i \mbox{\rm ~~for~} 1 \leq i < n ,
  \label{eqn:randomwalk}
\end{align}
in which the $\gamma_i \sim \normaldist(\mu, \sigma)$ are independent.
The random walk is \emph{unbiased} if $\mu = 0$ and \emph{biased} if $\mu \neq 0$.
We assume $\sigma = 1$ unless stated otherwise.

\smallskip
To evaluate the performance of local updates, we change the value of a single item, and since it depends on the amount, we fix parameters $\Delta \in \Rspace$ and $k \in \Nspace$ and do $2k+1$ parallel updates by changing the value of the $i$-th item from $\rwfunc (i)$ to $\rwfunc (i) + \delta_j$, with $\delta_j = \frac{j}{k} \Delta$ for all integers $j$ with $-k \leq j \leq k$.
For the experiments, we pick $\Delta = 5.0$  or $50.0$ and $k = 10$, while doing the update on a random walk of length between $10^2$ and $10^6$ generated for $\mu \in [-1,1]$ and $\sigma = 1$.
Each experiment is repeated one hundred times and averages as well as deviations from the average are reported.
For each experiment, we also measure the time to run \Gudhiname{} on the walk after the value change.
This provides the appropriate reference time, since \Gudhiname{} needs to recompute the persistence diagram from scratch after each update.

\smallskip
To evaluate the performance of topological updates, we pick a fraction, $c \in (0,1)$, generate a random walk of length $n$, and then cut it into a \emph{left walk} containing the first $\lfloor {c n} \rfloor$ items and a \emph{right walk} containing the remaining $\lceil {(1-c) n} \rceil$ items.
There are two such updates, which we now define.
To \emph{split}, we first construct the banana tree of the entire random walk, which we then cut into two  banana trees, one for the left and the other for the right walk.
To \emph{glue}, we first construct the banana trees of the left and right walks, which we then combine to a single banana tree for the walk of length $n$.
Doing this for fractions $c \in \{0.1, 0.3, 0.5, 0.7, 0.9\}$, we compare the time to split with the time used by \Gudhiname{} to construct the persistence diagrams of the left and right walks, and the time to glue with the time used by \Gudhiname{} to construct the persistence diagram of the entire random walk.

\subsection{Structural Properties}
\label{sec:3.2}\label{sec:random-walk-structure}

\begin{figure}[htb]
    \centering
    \includegraphics{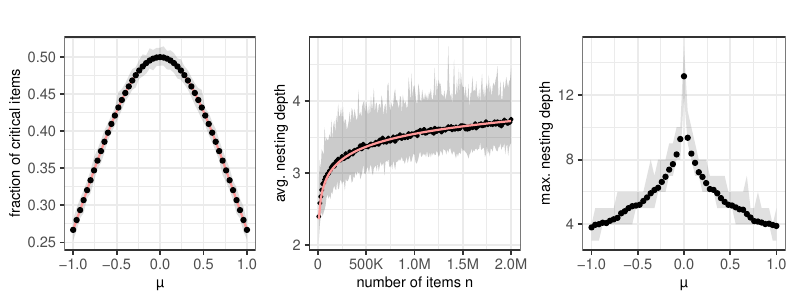}
    \caption{\footnotesize
        \emph{Left:} the average fraction of items that are critical as a function of the bias $\mu$.
        \emph{Middle:} the average nesting depth of leaf bananas in banana trees of unbiased random walks with $n$ items.
        The \emph{ribbon} extends from the minimum to the maximum observed value for each $n$; the dots mark the mean.
        The \emph{red line} is the graph of a constant times $\log n$ obtained by linear regression.
        \emph{Right:} the maximum nesting depth at $n=10^6$ items as a function of the bias $\mu$.
    }
    \label{fig:structure}
\end{figure}

We focus on three structural properties of banana trees, which have direct implications for the running time of our algorithms: the \emph{number of critical items}, the \emph{nesting depth of bananas}, and the \emph{lengths of trails}.
Since a banana tree does not store non-critical items, its number of nodes is the number of critical items.
In an unbiased random walk, the expected number of critical items is $50 \%$ of all items.
This fraction decreases when the bias increases, with about $27 \%$ for $\mu = \pm 1$; see the left panel in Figure~\ref{fig:structure} for more detailed information.

\begin{figure}
    \centering
    \includegraphics{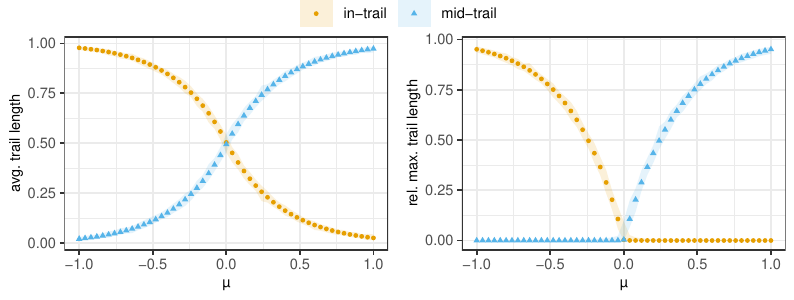}
    \caption{
        \emph{Left:} the average length of in-trails (\emph{orange dots}) and mid-trails (\emph{blue triangles}) in banana trees of random walks with bias $\mu$, averaged over all input sizes.
        \emph{Right:} the fraction of nodes on the longest in-trail (\emph{orange dots}) and longest mid-trails (\emph{blue triangles}) in banana trees of random walks with bias $\mu$, averaged over all input sized.
    }
    \label{fig:rw-trails}
\end{figure}

The middle panel of Figure~\ref{fig:structure} shows the average nesting depth of a leaf banana for an unbiased random walk, which appears to scale like the logarithm of the number of items.
In the unbiased case, even the maximum nesting depth seems to be small (about $13$ on average for a random walk of length $2 \cdot 10^6$), and scale logarithmically in the number of items.
With increasing bias, the maximum nesting depth decreases, to about $4$ at $\mu = \pm 1$; see the right panel of Figure~\ref{fig:structure}.
This is because in the biased case, there are fewer critical items but also the bananas tend to arrange in parallel rather than inside each other.
We observe that the nesting depth is about the same on average in up-trees and down-trees, both for biased and unbiased random walks, so our results hold for both trees.

\smallskip
For topological reasons, the mid-trails of the up-tree are the mid-trails of the down-tree, except that they are upside down.
In contrast, the in-trails of the two trees are not directly related because the in-panels in one direction become the out-panels in the other.
Notwithstanding, we observe that the length of the in-trails in the up-tree and the mid-trails in the down tree are very similar on the average, and by symmetry the same holds for the mid-trails in the up-tree and the in-trails in the down-tree.
These two facts do not contradict each other, because for biased data most of the critical items belong to the root banana, which does not adhere to the topological constraints just mentioned.
Let us therefore focus on the up-tree.
The left panel in Figure~\ref{fig:rw-trails} shows how the average length of in- and mid-trails depends on the bias.
In the unbiased case, both types of trails have the same length on average.
For positive $\mu$, the mid-trails tend to be longer than the in-trails, while the reverse happens for negative $\mu$.
The main cause for this trend is the longest trail, which starts to dominate the others in length as the bias increases;
see the right panel in Figure~\ref{fig:rw-trails}.
Assuming $\mu > 0$, the global minimum and maximum tend to lie to the left and right of the middle, respectively, with the majority of the items between them.
By convention, the connecting in-trail and mid-trail contain the items to the left and right of the global minimum, respectively.
This explains why the mid-trail dominates in this case, and why the in-trail dominates when $\mu < 0$.
Indeed, for $\mu = \pm 1$, we observe about $95 \%$ of the internal nodes (the local maxima) on the longest trail.

\subsection{Local Maintenance}
\label{sec:3.3}

We evaluate the performance of banana trees when the value of a single item changes by $\delta$.
Note that the corresponding update is a combination of some number of interchanges and at most one cancellation, at most one anti-cancellation, and at most two slides \cite{CEHO24}.
Not surprisingly, the time increases with the amount of change.
For example, there is no effect at all on the persistence diagram for about $70 \%$ of the updates with $\delta = \pm 0.26$, for about $36 \%$ of the updates with $\delta = \pm 0.79$, but only about $1 \%$ for updates with $\delta = \pm 3.4$.
Taking this into account, it is not surprising that banana trees are faster than \Gudhiname{} when $\delta = \pm 0.26$ for all lengths of random walks we tried.
\begin{figure}[htb]
    \centering
    \includegraphics{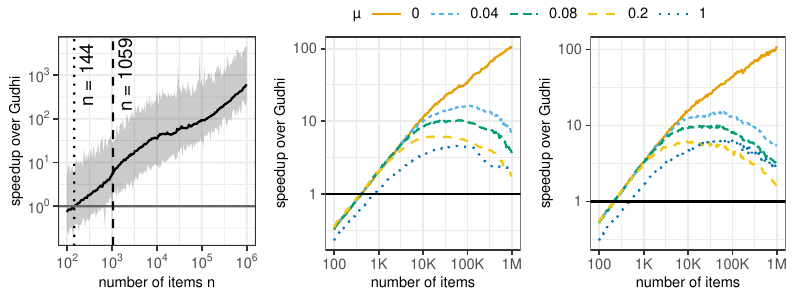}
    \caption{\footnotesize
      Comparing the maintenance of banana trees with reconstructing the persistence diagram with \Gudhiname{}, in which the baseline of no speedup ($10^0 = 1$) is marked with a \emph{horizontal gray} line.
      \emph{Left:} the speedup for updating a value with $\delta = \pm 5.0$ depending on the length of the random walk, $n$.
      The ribbon spans the minimum and maximum observed speedup, with the \emph{black curve} tracing the median speedup.
      \emph{Middle:} the speedup for using banana trees to cut a random walk with bias $\mu$ in half.
      The type and color of the curve encodes the amount of bias, and each curve shows the median speedup over a hundred repeats for each $n$.
      \emph{Right:} the speedup for using banana trees to concatenate two equally long random walks with bias $\mu$.}
    \label{fig:maintenance-speedup}
\end{figure}

The left panel in Figure~\ref{fig:maintenance-speedup} focuses on the case $\delta = \pm 5.0$.
As indicated by the vertical dashed and dotted lines, banana trees are faster than \Gudhiname{} for all random walks of length $n \geq 1059$, and faster in the median for all $n \geq 144$.
Not shown in the panel is that the speedup decreases with increasing bias.
In particular, for $n = 10^6$, the speedup of $612$ at $\mu = 0$ decreases to $258$ at $\mu = 1$, and for $n = 1059$, the speedup of $6.0$ at $\mu = 0$ decreases to $2.9$ at $\mu = 1$.
This can be explained by recalling that for large bias the majority of the internal nodes belong to the longest trail, which connects the global minimum to the global maximum.
Hence, items with similar values are likely to be near each other and thus require interchanges when an update is performed.
However, even for a large change, such as for $\delta = \pm 50.0$, maintaining the banana tree is still faster than reconstructing the persistence diagram with \Gudhiname{}.
For example, the median speedup for $n = 10^4$ exceeds $10$ and for $n = 10^6$ it exceeds $100$.

\subsection{Topological Maintenance}
\label{sec:3.4}

We call the operations of \emph{cutting} a list into two, and \emph{concatenating} two lists to one topological because they change the number of lists in the overall organization of the data.
We begin with evaluating the performance of cutting a random walk.
The middle panel in Figure~\ref{fig:maintenance-speedup} shows the speedup over \Gudhiname{} when we cut a biased or unbiased random walk in half.
In the unbiased case, the speedup increases with the length of the walk.
Recalling the discussion of structural properties, this can be rationalized by noticing that the maximum nesting depth bounds the number of bananas that need to be split, and the trail length bounds the time to reorganize bananas.
The maximum nesting depth scales like $\log n$ and the average trail length is only $0.5$, so we anticipate logarithmic time for splitting, which we indeed observe in our experiments.
Altering the position of the cut increases the speedup, namely by about $2 \%$ for $c = 0.5 \pm 0.2$ and by about $7 \%$ for $c = 0.5 \pm 0.4$.

\smallskip
The advantage of the banana tree over \Gudhiname{} diminishes when the random walks get progressively more biased.
The reason is again that the larger the bias, the larger the fraction of internal nodes in the longest trail of the banana tree.
Splitting the corresponding banana requires the resetting of many pointers stored in nodes along this trail,
which in some cases costs time proportional to the number of critical items, which we indeed observe in our experiments.
Unfortunately, this more than compensates for the low nesting depth, which guarantees that only a small number of bananas need to be split.
The upper half of Table~\ref{tbl:topological-times} shows the running time and the speedup over \Gudhiname{} for cutting a random walk in half.
\begin{table}[htb]
    \centering
    \caption{\footnotesize \emph{Upper half}: the average time for cutting a random walk of length $n = 10^6$ in half, and the speedup over \Gudhiname{}.
    \emph{Lower half}; the average time for concatenating two random walks of length $n = 10^6/2$ each, and the speedup over \Gudhiname{}.}
    \footnotesize \vspace{0.05in}
    \begin{tabular}{lr||rrrrr}
                   & &$\mu=0$& $\mu=0.04$   & $\mu=0.08$   & $\mu=0.2$    & $\mu=1$ \\
        \midrule \midrule
        Cut  & time in $\mu\mathrm{s}$ & $93$  & $1417$ & $2807$ & $5350$ & $1683$ \\ 
          & speedup & $105.00$ & $7.24$ & $3.54$ & $1.68$ & $2.15$ \\ \midrule
        Concatenate & time in $\mu\mathrm{s}$ & $90$  & $1809$ & $3041$ & $5138$ & $1228$ \\         
          & speedup & $105.00$ & $5.31$ & $3.12$ & $1.74$ & $3.22$
    \end{tabular}
    \label{tbl:topological-times}
\end{table}

We finally address the reverse operation, that of concatenating two random walks or, correspondingly, of gluing two banana trees.
The curves in the right panel of Figure~\ref{fig:maintenance-speedup} are similar to those in the middle panel, which suggests that the performance is similar to that of cutting, which is confirmed by the numbers in Table~\ref{tbl:topological-times}.
For $n \leq 10^4$, gluing appears to be slightly faster than splitting, but the difference is less clear already for $n = 10^6$.

\section{Special Time Series}
\label{sec:4}

This section considers three types of time series:  \emph{worst-case examples} to expose when banana trees under-perform, \emph{quasi-periodic data} to illustrate how banana trees reflect periodicity, and \emph{real-world data} to demonstrate that banana trees are not just theory.
While banana trees struggle for data of the first type, they mirror the performance observed in unbiased random walks for the latter two types.

\subsection{Worst-Case Examples}
\label{sec:4.1}

The size of the augmented persistence diagram (number of points and arrows) is proportional to the number of critical items.
There are configurations in which a single update changes almost the entire diagram and thus takes time at least linear in the number of such items.
We construct such \emph{worst-case inputs} and study the performance of banana trees when confronted with such data.
\begin{figure}[htb]
    \centering
    \includegraphics{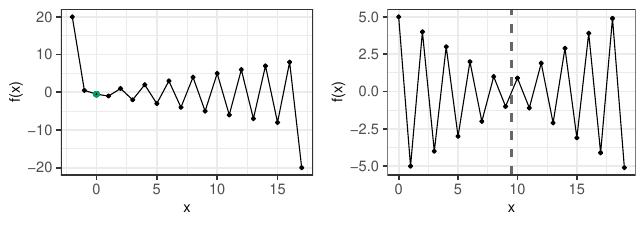}
    \caption{\footnotesize
      Worst-case examples for local and topological maintenance.
      \emph{Left:} to increase the value of the marked item triggers a linear number of interchanges.
      \emph{Right:} to cut the list at the \emph{dashed line} affects every persistent pair.
      Both operations take time linear in the number of critical items, which for the two time series are all or almost all items.}
    \label{fig:linear-case-inputs}
\end{figure}

Consider first the operation that increases the value of the marked item in the left panel of Figure~\ref{fig:linear-case-inputs}.
Initially, we have a linear sequence of double-panel windows, each nested inside the next.
When the value of the marked item passes that of its left neighbor, an anti-cancellation adds the banana they span to the banana tree, and its window is nested inside the others.
As we increase the value further, the marked item interchanges with the maxima to its right, each time moving up one level within the sequence of windows.
When the marked item finally becomes the global maximum, it will have interchanged with all other maxima, which takes $\Theta(n)$ time.
For $n = 10^2$ and $n = 10^6$, we measured $3.4 \mu {\rm s}$ and $4.7 {\rm m} {\rm s}$ on average, respectively.
This corresponds to a median speedup by a factor of $0.05$ and $0.02$ if compared with \Gudhiname{}, which is really a slowdown by a factor of $20$ and $50$, respectively.

\begin{table}[hbt]
    \centering
    \caption{\footnotesize
      Performance of splitting and gluing banana trees for the example time series in the right panel of Figure~\ref{fig:linear-case-inputs}, showing the median slowdown/speedup if compared to \Gudhiname{}.
      The parameter interpolates between these time series at $\lambda = 0$ and unbiased random walks at $\lambda = 1$.}
      \vspace{0.1in}
      \footnotesize
    \begin{tabular}{l||rrrr|rrrr}
        & \multicolumn{4}{c|}{split} & \multicolumn{4}{c}{glue} \\
                   & $\lambda = 0.0$ & $0.0001$  & $0.01$ & $1.0$  & $\lambda = 0.0$ & $0.0001$  & $0.01$   & $1.0$  \\
        \midrule\midrule
        $n = 10^2$ & 0.06            & 0.06      & 0.04   & 0.35   & 0.40            & 0.44      & 0.68     & 2.38   \\
        $n = 10^4$ & 0.04            & 0.03      & 5.81   & 11.20  & 0.46            & 0.83      & 33.6     & 41.40  \\
        $n = 10^6$ & 0.09            & 4.71      & 54.70  & 111.00 & 0.46            & 81.0      & 255      & 397.00 
    \end{tabular}
    \label{tbl:topological-wc-speedup}
\end{table}
Consider second the operation that cuts the time series in the right panel of Figure~\ref{fig:linear-case-inputs} in the middle, which is marked by the vertical dashed line.
We observe an average running time between $10 \mu {\rm s}$ at $n = 10^2$ and $72 {\rm m} {\rm s}$ at $n = 10^6$.
Compare this with an average running time of $7 \mu {\rm s}$ at $n = 10^2$ and $48 {\rm m} {\rm s}$ at $n = 10^6$ for concatenating the two lists back to the original length.
This comparison agrees with the earlier observation that gluing banana trees is slightly faster than splitting them.
Table~\ref{tbl:topological-wc-speedup} lists the speedup if compared to \Gudhiname{}, which for a factor less than $1$ is really a slowdown.
To create a more informative experiment, we interpolate between these time series and unbiased walks, with drastic improvements even for small $\lambda > 0$; see Table~\ref{tbl:topological-wc-speedup}.
Indeed, the noise quickly flattens the banana tree by decreasing the length of trails, which explains the improvement.

\subsection{Quasi-Periodic Data}
\label{sec:4.2}

Random walks are not periodic, but real-world time series often exhibit some amount of periodicity, at least approximately.
To study banana trees under these conditions, we construct what we call \emph{quasi-periodic} inputs.
This term does not have a mathematical definition, and for our experiments just means a periodic signal that is randomly perturbed.
Specifically, we generate such input by modifying the iterative construction in \eqref{eqn:randomwalk}:
\begin{align}
  \rwfunc (0) = 0 \mbox{\rm ~~and~~} r(i) = r(i-1) + \eta_i \mbox{\rm ~~for~} 1 \leq i < n ,
  \label{eqn:quasirandomwalk}
\end{align}
in which the $\eta_i \sim \normaldist(\mu_i, \sigma)$ are independent and normally distributed, with the mean changing periodically, $\eta_i = \sin (2 \pi \omega i)$, as governed by the \emph{frequency parameter} $0 \leq \omega < 1$, which we fix to $\omega = 5/n$.
It will be useful to compare the influence of the standard deviation, so we no longer assume $\sigma = 1$.
\begin{figure}[hbt]
    \centering
    \includegraphics{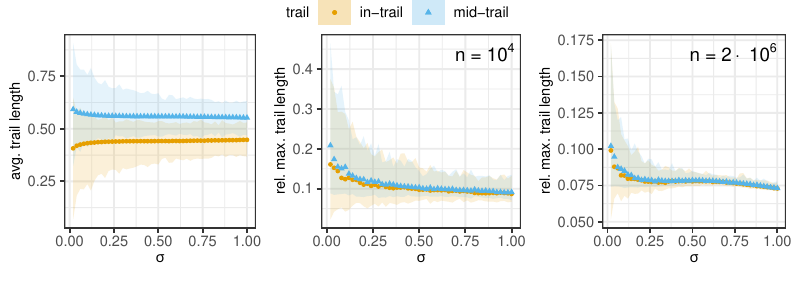}
    \vspace{-0.25in}
    \caption{\footnotesize
        Measuring trail lengths.
        The in-trails and mid-trails are marked with \emph{orange dots} and \emph{blue triangles}, respectively, and the ribbons extend from the minimum to the maximum observed values.
        \emph{Left:} the average length of in-trails and mid-trails depending on the standard deviation and averaged over all input sizes.
        \emph{Middle and right:} the fraction of nodes on the longest in-trail and mid-trail, again depending on the standard deviation but now averaged over inputs of size $10^4$ and $2 \cdot 10^6$, respectively.}
    \label{fig:quasitrails}
\end{figure}

For large $\sigma$, the banana trees of the time series show the features we have seen for unbiased random walks, while for small $\sigma$, they locally behave like biased random walks.
For example, for $\sigma = 1$, we observe about $37 \%$ of the items to be critical, while for $\sigma = 0.5$ and $\sigma = 0.02$ only about $21 \%$ and $1 \%$ of the items are critical, on average.
Like the number of critical items, also the maximum nesting depth for quasi-periodic signals is similar to that of random walks:  it scales like $\log n$ for large $\sigma$ and decreases as $\sigma$ goes to $0$.
In contrast, the average nesting depth is independent of $n$, ranging from $2$ to $4.7$ with mean $2.4$.
The behavior of the trail length is illustrated in Figure~\ref{fig:quasitrails}.
As shown in the left panel, the average trail length is about $0.5$, with consistently shorter in-trails than mid-trails, on average.
This difference is more pronounced for small $\sigma$, when the quasi-periodic signal behaves more like a biased rather than unbiased random walk.
The middle and right panels show how the fraction of internal nodes on the longest trail increases as $\sigma$ goes to $0$.

\smallskip
We conclude that banana trees on quasi-periodic signals with large standard deviation resemble unbiased random walks in terms of their structural parameters, and drift toward biased random walks as the standard deviation goes to $0$.

\subsection{Real World Examples}
\label{sec:4.3}\label{sec:real-world-examples}

We analyze structural properties of banana trees constructed on three types of real-world time series:
\medskip \begin{itemize}
  \item electrocardiography (ECG) data from PhysioNet~\cite{GAG00,PGS20,RSG21}; limiting ourselves to recordings of length at least $10^5$, each separated into up to 12 leads, gives 32443 time series in total;
  \item audio data from EasyCom~\cite{DTL21},
  of which we use 1336 audio files;
  \item physical activity data from PAMAP2~\cite{Rei21} (temperature, heart rate along with recordings from accelerometers, gyroscopes, and magnetometers), of which we use 560 time series.
\end{itemize} \medskip
We preprocess the data by adding uniformly distributed noise to ensure all values are distinct, making sure the noise is small enough so it breaks ties but does not otherwise alter the ordering.
\begin{table}[hbt]
    \centering
    \caption{\footnotesize
      Structural properties of the banana trees constructed for time series from three databases.
      The average, minimum, and maximum of the maximum nesting depth is measured relative to the depth we observe for unbiased random walks with the same fraction of critical items.}
      \vspace{0.00in}
      \footnotesize
    \begin{tabular}{l||c|ccc|cc}
        & \multicolumn{1}{c|}{fraction of} & \multicolumn{3}{c|}{max nesting depth} & \multicolumn{2}{c}{average length}  \\
        & \multicolumn{1}{c|}{critical items} & avg & min & max & in-trail  & mid-trail  \\ \midrule \midrule
      PhysioNet & 0.24 & 1.2 & 0.8 & 6.3 & 0.48 & 0.52  \\
      EasyCom   & 0.22 & 2.7 & 1.6 & 4.9 & 0.48 & 0.52  \\
      PAMAP2    & 0.48 & 1.9 & 1.2 & 3.2 & 0.49 & 0.51
    \end{tabular}
    \label{tbl:real-world-data}
\end{table}

We observe that the average max nesting depth for the three data sets is only a small factor larger than for unbiased random walks; see the middle three columns in Table~\ref{tbl:real-world-data}.
The largest maximum nesting depth for PhysioNet data is a bit higher, by a factor $6.3$, but $99 \%$ of the time series from this data have maximum nesting depth at most $1.68$ times that for unbiased random walks.
The average trail lengths are again similar to those of unbiased random walks---see the last two columns, with standard deviation about $0.05$ throughout---and so are the fractions of items in the longest trails, whose geometric means are $0.8\%, 0.008\%, 0.1\%$ of all critical items for the in-trails and $0.8\%, 0.008\%, 0.2\%$ of all critical items for the mid-trails in the three data sets, compared with about $0.3\%$ and $0.4\%$ in unbiased random walks.

\section{Discussion}
\label{sec:5}

We provide an implementation of the banana tree data structure and supply experimental evidence demonstrating its efficiency. 
The work reported in this paper confirms the theoretical advantages of banana trees over static alternatives and opens up avenues for further theoretical inquiries and practical applications.
One intriguing question that arises in the context of quasi-periodic data is whether we can determine the (possibly varying) period of the data without prior knowledge, or provide meaningful quantification of the extent to which a signal deviates from being periodic. 

\smallskip
The potential applications of banana trees extend beyond academic research into practical, real-world scenarios. 
Comparisons with other methods and identifying impactful use cases represents a promising next step to provide valuable insights in fields such as healthcare, where real-time analysis of patient data is crucial, or finance, where the swift analysis of market data is essential. 

\bibliography{bibliography}

\clearpage
\appendix

\section{Maintaining Banana Trees}
\label{app:A}

\begin{figure}[hpt]
    \centering
    \sidecaption{fig:windows-and-bananas-2}
    \raisebox{-\height}{
        \tikzmath{ \windowbananascale=0.45; }
        \input{graphics/tikz/windows}
        \input{graphics/tikz/bananas}
    }
    \sidecaption{fig:banana-ptrs}
    \raisebox{-\height}{
        \tikzmath{ \windowbananascale=0.4;}
        \input{graphics/tikz/banana-pointers}
    }
    \caption{
        \refinsubcaption{fig:windows-and-bananas-2}
            A window spanned by minimum $u$ and maximum $p$ on the left and the corresponding section of the banana tree.
            Rectangles indicate windows with the in-panel and mid-panel separated by a dotted line.
        \refinsubcaption{fig:banana-ptrs}
            An internal node $a$ and a leaf node $b$ that span a banana $\banana{b}{a}$ and pointers defined at these nodes.
            The pointer $\lowptr(b) = b$ is not shown.
    }
    \label{fig:banana-tree-implementation}
\end{figure}
Banana trees are described by pointers between nodes, as illustrated in \cref{fig:banana-ptrs}.
Each node has pointers \inptr{} and \midptr{}, which point to one end of the doubly-linked lists that represent the in-trail and mid-trail.
The nodes on a trail are connected by pointers \upptr{} and \dnptr{}, which point to the next node with higher and lower value, respectively.
Each minimum stores a pointer \dthptr{}, which points to the maximum that it is paired with.
Note that if $\dthptr{b} = a$, then nodes $b$ and $a$ are the ends of an in-trail and a mid-trail.
This establishes the pairing in the persistence diagram.
Finally, every node stores a pointer \lowptr{} that points to the lower end of the trail containing the node, or the node itself if it is a minimum.
In the example in \cref{fig:windows-and-bananas-2}, $\lowptr(r) = \lowptr(t) = b$.
The pointer \lowptr{} can be used to compute the inverse of the pointer \dthptr{}.
For example, given $\dthptr(d) = c$ we have $\lowptr(\inptr(c)) = \lowptr(\midptr(c)) = d$.
Furthermore, every node stores a pointer to the item it represents.

\begin{figure}
    \centering
    \sidecaption{fig:anticancellation-function}
    \raisebox{-\height}{
        \tikzmath{ \actikzscale=0.5; }
        \input{graphics/tikz/anticancellation-function}
    }
    \sidecaption{fig:anticancellation-tree}
    \raisebox{-\height}{
        \tikzmath{ \actikzscale=0.5;}
        \input{graphics/tikz/anticancellation-tree}
    }
    \caption{
        \refinsubcaption{fig:anticancellation-function} Increasing the value of a non-critical item $j$ above that of a neighboring non-critical item $k$ results in a new window spanned by the minimum $k$ and the maximum $j$.
        \refinsubcaption{fig:anticancellation-tree} To update the banana tree, we start at the maximum $m$ closest to $j$ such that $m < k < j$
            and, beginning with $\midptr(m) = w$ and following $\inptr(\cdot)$-pointers (dotted red arrows), we search the last node with value greater than $f(j)$; here we land at the node $u$.
            Then a new banana $\banana{k}{j}$ is inserted,
            in this case with $\inptr(u)$ being set to point to $j$.
    }
    \label{fig:anticancellation}
\end{figure}
Banana trees can be constructed in linear time by traversing the critical items from left to right.
During the traversal, each item is processed once to be inserted into the banana tree
and once to assign pointers representing the nesting of bananas and the pairing of items.
We refer to \cite{CEHO24} for details and focus on the maintenance algorithms.

We first discuss how the banana trees are updated when the value of a single item changes.
An arbitrary value change is processed by a sequence of \emph{local maintenance operations}, which we describe here.
Let $f: [1,m] \to \Rspace$ and $g: [1,m] \to \Rspace$ be the maps before and after the update, respectively.
We assume that the value of item $j$ is changed and $f(j) \neq g(j)$.
The maps $f$ and $g$ agree in the values of all other items, i.e., $f(i) = g(i)$ for all $i \neq j$.

If $j$ is non-critical in $f$, it may become critical in $g$.
We distinguish two cases:
case (1) where $j$ becomes critical by passing the value of a critical neighbor, i.e., by its value either increasing above a neighboring maximum or decreasing below a neighboring minimum,
and case (2) where $j$ becomes critical by passing the value of a non-critical neighbor.
In case (1) the banana tree is updated by a \emph{slide}: the critical neighbor becomes non-critical, and thus its node in the banana tree is simply updated to refer to $j$.
In case (2) the item $j$ and its neighbor $k$ both become critical and $j$ either becomes a maximum or a minimum, depending on whether its value increased or decreased.
Assume $j$ becomes a maximum as illustrated in \cref{fig:anticancellation}.
The banana tree is updated by inserting a new banana $\banana{k}{j}$.
To find where to insert the new banana the algorithm begins at the maximum $m$ such that $m < k < j$ or $j < k < m$.
If $j < m$ it follows the leftmost path beginning at the child of $m$ immediately to its right;
if $j > m$ it follows the rightmost path beginning at the child of $m$ immediately to its left.
The node $j$ is inserted below the last node $d$ such that $f(d) > f(j)$.
See the dotted path in \cref{fig:anticancellation}.

If $j$ is critical in $f$, it may become non-critical in $g$.
It may either (a) become non-critical by passing the value of a non-critical neighbor---which is the reverse of case (1) above---or
(b) become non-critical by passing the value of a critical neighbor---which is the reverse of case (2) above.
In case (a) the banana tree is also updated by a \emph{slide} (see above);
in case (b) the critical neighbor of $j$ also becomes non-critical.
They will span a banana when this happens and we only need to only this banana to update the banana tree.
This operation called a \emph{cancellation}.

\begin{figure}
    \centering
    \sidecaption{fig:max-interchange}
    \tikzmath{ \maxxchangetikzscale=0.45; }
    \raisebox{-\height}{
        \input{graphics/tikz/max-interchange-1}
    }
    \raisebox{-\height}{
        \input{graphics/tikz/max-interchange-2}
    }
    \sidecaption{fig:min-interchange}
    \raisebox{-\height}{
        \input{graphics/tikz/min-interchange}
    }
    \caption{\refinsubcaption{fig:max-interchange} An interchange of maxima where the value of $j$ increases above that of $q$.
            Red trails are in-trails and black trails are mid-trails unless labeled otherwise.
            In the right case, the in-trail of $\banana{i}{j}$ becomes the mid-trail of $\banana{i}{q}$,
            and the mid-trail of $\banana{i}{j}$ becomes the in-trail of $\banana{i}{q}$.
            There are two additional cases that are similar to the reverse of the cases shown here.
        \refinsubcaption{fig:min-interchange} An interchange of minima where the value of $i$ decreases below the value of $p$.
            The nodes $a$, $b$, and $j$ satisfy $f(a) < f(j) < f(b)$.
            For nodes in sections of trail between $j$ and $q$ the $\lowptr$ pointer is updated to point to $i$.
        }
    \label{fig:interchanges}
\end{figure}
Finally, an item $j$ that is critical may remain critical in $g$, but changing its value may affect the order of critical items.
We first consider the case where $j$ is a maximum.
The banana tree only needs to be updated if the value of $j$ increases above $\upptr(j)$, or if it decreases below $\midptr(j)$, $\inptr(j)$ or $\dnptr(j)$ (whichever has greatest value).
In this case we perform an \emph{interchange of maxima}, which resembles a rotation and which is illustrated in \cref{fig:max-interchange}.
Note that this operation only reassigns a constant number of pointers.
The second case to consider is the one where $j$ is a minimum.
The banana tree needs to be updated if either the value of $j$ falls below the value of $\lowptr(\dthptr(j))$, or rises above the value of a node $k$ with $j = \lowptr(\dthptr(k))$.
This is achieved by an \emph{interchange of minima}, which is illustrated in \cref{fig:min-interchange}.

Finding the nodes with which to perform an interchange of minima may be expensive, especially in the case where the value of $j$ increases.
We avoid this by exploiting the symmetry between updates in the up-tree and the down-tree:
an interchange of minima in the up-tree occurs only if there is a corresponding interchange of maxima in the down-tree involving the same items,
and an interchange of minima in the down-tree only occurs together with a corresponding interchange of maxima in the up-tree.
When changing the value of a minimum $j$, the update is performed in the down-tree, where $j$ is a maximum.

Cancellations and slides take time $O(\log n)$, since they require items to be moved between the binary search trees.
An interchange of maxima takes constant time; an interchange of minima takes time $O(k)$, where $k$ is the number of changes to the augmented persistence diagram due to the interchange.
Finally, anti-cancellations take time $O(\log n + k')$, where $k'$ is upper bounded by the nesting depth of the inserted banana.

\begin{figure}
    \centering
    \tikzmath{ \topotikzscale = 0.5; }
    \sidecaption{fig:injury}
    \raisebox{-\height}{
        \input{graphics/tikz/injury}
    }
    \hspace{0.5cm}
    \sidecaption{fig:fatality}
    \raisebox{-\height}{
        \input{graphics/tikz/fatality}
    }
    \caption{
        \refinsubcaption{fig:injury} Processing an injury for cutting.
            The green dashed section of the in-trail of $\banana{j}{k}$ is inserted into the mid-trail above $a$.
        \refinsubcaption{fig:fatality} Processing a fatality for cutting.
            The green dashed section of the banana $\banana{j}{k}$ is swapped with $a$.
            This section consists of the in-trail of $\banana{j}{k}$, the node $j$ and a section of the mid-trail of $\banana{j}{k}$.
            After the swap the in-trail between $a$ and $k$ is empty, shown as a dotted line.}
    \label{fig:topological-operations}
\end{figure}
We now discuss how banana trees are updated when splitting an interval or concatenating two intervals.
There are two corresponding \emph{topological maintenance operations}, which \emph{cut} and \emph{glue} the banana trees.

We first describe how to cut a banana tree between items $j$ and $j+1$ at a virtual item $c = j + \frac{1}{2}$.
The cases $f(j) > f(j+1)$ and $f(j) < f(j+1)$ are symmetric and we assume the former.
First, items $p$ and $q$ are inserted such that $j < p < q < j+1$ and $f(j) > f(q) > f(p) > f(j+1)$,
setting $f(q) = f(p) + \eps$ with $\eps > 0$ such that $p$ and $q$ are paired.
The items $p$ and $q$ become the new endpoints of the left and right interval, respectively,
and are inserted into the banana trees via an anti-cancellation.
The bananas where $c$ lies in one of the three panels of the corresponding window are stored in a stack,
with $\banana{p}{q}$ on the bottom and the value of the maximum increasing towards the top of the stack.
Each banana $\banana{a}{b}$ on the stack is classified according to the position of $c$ in relation to $a$ and $b$:
if $c < a < b$ or $b < a < c$, then $c$ lies in the in-panel, and we say that $\banana{a}{b}$ \emph{experiences an injury};
if $a < c < b$ or $b < c < a$, then $c$ lies in the mid-panel, and we say that $\banana{a}{b}$ \emph{experiences a fatality};
if $a < b < c$ or $c < b < a$, then $c$ lies in the out-panel, and we say that $\banana{a}{b}$ \emph{experiences a scare}.
To perform actual cutting an empty tree is constructed consisting of a special root and a dummy node.
The bananas on the stack are then processed from top to bottom.
Based on injury, fatality or scare nodes are moved between the two trees,
until $\banana{p}{q}$ has been processed and the two trees correspond to the left and right list.
\Cref{fig:topological-operations} illustrates the processing of injuries and fatalities.
Scares are processed via an interchange of minima.

Gluing is the reverse of cutting, but does not require a stack.
By following the rightmost path of the left tree and the leftmost path on the right tree upwards from the endpoints
the minima of short-wave windows are traversed by decreasing value.
At each step, based on the values of unprocessed maxima, the algorithm either reverses an injury, a fatality or a scare,
undoing the operations illustrated in \cref{fig:topological-operations}.
The algorithm ends when one of the trees is empty.

The running time for both algorithms is $O(\log n + k)$, where the logarithmic term comes from splitting or joining the binary search trees
and $k$ is the number of changes to the augmented persistence diagram.

\section{Additional Experiments}

\FloatBarrier
\subsection{Random Walks}
\begin{figure}[ht]
    \centering
    \begin{subfigure}{0\linewidth}\phantomcaption\label{fig:rw-construction-unbiased}\end{subfigure}
    \begin{subfigure}{0\linewidth}\phantomcaption\label{fig:rw-construction-alloc}\end{subfigure}
    \begin{subfigure}{0\linewidth}\phantomcaption\label{fig:rw-construction-biased}\end{subfigure}
    \includegraphics{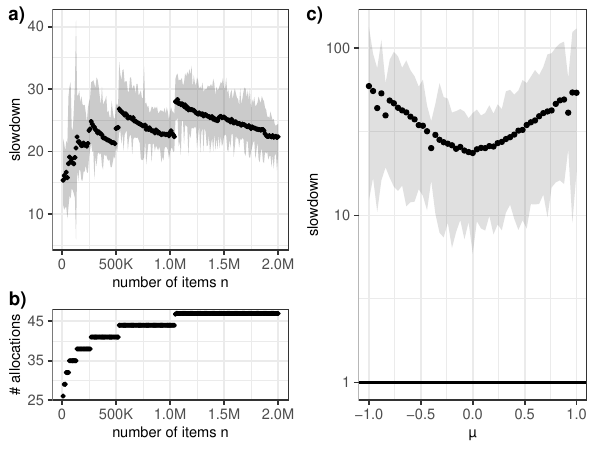}
    \caption{Timing the construction of banana trees.
    \refinsubcaption{fig:rw-construction-unbiased}
        Slowdown of banana trees compared to \Gudhiname{} for \emph{unbiased} random walks ($\mu = 0$).
        For each number of items, $n$, the ribbon extends from the minimum to the maximum observed slowdown, and the point shows the geometric mean.
    \refinsubcaption{fig:rw-construction-alloc}
        Average number of allocations by the memory pools for storing banana tree nodes and items.
    \refinsubcaption{fig:rw-construction-biased}
        Slowdown compared to \Gudhiname{} for different values of the parameter $\mu$ in the generation of random walks.
        The ribbon extends from the minimum to the maximum observed slowdown, and the point shows the geometric mean.}
    \label{fig:random-walk-construction}
\end{figure}
\Cref{fig:rw-construction-unbiased} shows the slowdown over \Gudhiname{} for constructing banana trees on unbiased random walks.
Average slowdowns range from $15.4$ to $28.3$.
There are jumps in the slowdown of banana tree construction at exponentially spaced intervals.
These are due to allocations in the memory pools used for storing items and nodes.
Note how the increases in allocations are aligned with the jumps in slowdown.
On biased random walks \Gudhiname{} is even faster relative to the banana trees.
The average slowdown increases to $60$ for $\mu = -1$ and to $54$ for $\mu = 1$, on average over all input sizes.

\begin{figure}[hbtp]
    \centering
    \begin{subfigure}{0\linewidth}\phantomcaption\label{fig:rw-maxnesting-vs-n}\end{subfigure}
    \begin{subfigure}{0\linewidth}\phantomcaption\label{fig:rw-maxnesting-vs-mu}\end{subfigure}
    \includegraphics{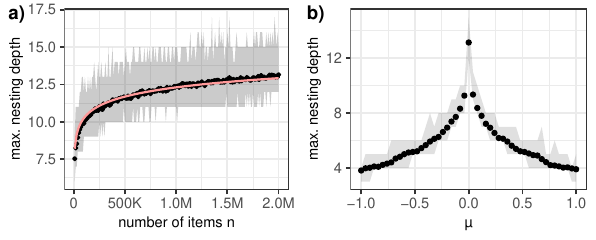}
    \caption{
        \refinsubcaption{fig:rw-maxnesting-vs-n}
            The maximum nesting depth of the leaf bananas in the up-tree of an \emph{unbiased} random walk ($\mu = 0$) of size $n$.
            The ribbon extends from the minimum to the maximum obsered average nesting depth, and the dot shows the mean.
            The red line is the graph of a function of the form $\beta_1 \log n$ fit via linear regression.
        \refinsubcaption{fig:rw-maxnesting-vs-mu}
            Maximum nesting depth of 100 up-trees for \emph{biased} random walks of size $2\cdot10^6$.
            The ribbon extends from the minimum to the maximum per bias parameter, $\mu$, and the dot shows the mean.
    }
    \label{fig:random-walk-maxnesting}
\end{figure}

\Cref{fig:rw-maxnesting-vs-n} shows how the maximum nesting depth on unbiased random walks grows with input size.
Note the logarithmic scaling as discussed in \cref{sec:random-walk-structure}.
The maximum nesting depth decreases as the random walks become more biased.
This is shown for random walks of length $2\cdot10^6$ in \cref{fig:rw-maxnesting-vs-mu}.

\FloatBarrier
\subsection{Linear-Time Cases}

\subsubsection{Definitions}
We now define the functions underlying the worst-case examples displayed in \cref{fig:linear-case-inputs}.
For local maintenance we define a function $\wclmfunc: \Zspace \to \Rspace$ over $n$ items for some even $n > 5$ as
\begin{equation}
    \label{eq:local-worst-case}
    \begin{aligned}
        \wclmfunc(-2)    &= n, \quad \wclmfunc(-1) = 0.5, \quad \wclmfunc(0) = -0.5, \quad \wclmfunc(n - 3) &= -n \\
        \wclmfunc(i)     &= (-1)^i \cdot \flfrac{i+1}{2}  \quad \text{for } 0 < i < n - 3. 
    \end{aligned}
\end{equation}
We then consider two value changes: (1) changing $\wclmfunc(0)$ to $0.5 + \eps$ for some $\eps \in (0, 1.5)$ and (2) changing $f(0)$ to $n + \eps$ for some $\eps > 0$.
Change (1) causes an anti-cancellation that takes linear time; change (2) additionally leads to $\Theta(n)$ interchanges.

The linear-time case for topological maintenance is a function $\wctmfunc: \Zspace \to \Rspace$ defined over $n$ items, with $n$ a multiple of $4$, which we give below.
First, we define an auxiliary function
\begin{equation}
    \label{eq:topological-worst-case-stage}
    s(i) = (-1)^i \cdot \left(\flfrac{i}{2} + 1\right).
\end{equation}
The function $\wctmfunc$ consists of two stages derived from $s$:
\begin{equation}
    \label{eq:topological-worst-case}
    \begin{aligned}
        &\wctmfunc(i)    &= -s\left(\frac{n}{2} - 1 - i\right)       && \qquad\text{for } 0 \leq i < \frac{n}{2},   \\
        &\wctmfunc(i)    &= s\left(i - \frac{n}{2}\right) - 0.1      && \qquad\text{for } \frac{n}{2} \leq i < n.
    \end{aligned}
\end{equation}
Cutting a banana tree over this function involves $O(n)$ bananas; similarly, gluing the sub-intervals given in \eqref{eq:topological-worst-case} requires processing $O(n)$ bananas.

We also study how deviating from the linear-time cases impacts the running time.
To this end we interpolate between $\wclmfunc$ ($\wctmfunc$) and an unbiased random walk, controlled by parameter $\lambda \in [0,1]$:
\begin{align}
    \wclmfunc(i;\lambda) &= (1 - \lambda) \cdot \wclmfunc(i)  + \lambda \cdot \max(\wclmfunc(i)) \cdot r(i;0,1),  \label{eq:linear-wc-lerp}  \\
    \wctmfunc(i;\lambda) &= (1 - \lambda) \cdot \wctmfunc(i)  + \lambda \cdot \max(\wctmfunc(i)) \cdot r(i;0,1),  \label{eq:topological-wc-lerp}
\end{align}
where we scale the random walk by the maximum function value of the linear-time case.
At $\lambda = 0$ we have the pure linear-time with no noise; as $\lambda$ increases towards $1$ the function is transformed into an unbiased random walk.

\subsubsection{Additional Results}
\begin{figure}[ht]
    \centering
    \begin{subfigure}{0\linewidth}\phantomcaption\label{fig:local-wc-speedup}\end{subfigure}
    \begin{subfigure}{0\linewidth}\phantomcaption\label{fig:cut-wc-speedup}\end{subfigure}
    \begin{subfigure}{0\linewidth}\phantomcaption\label{fig:glue-wc-speedup}\end{subfigure}
    \includegraphics[width=0.9\linewidth]{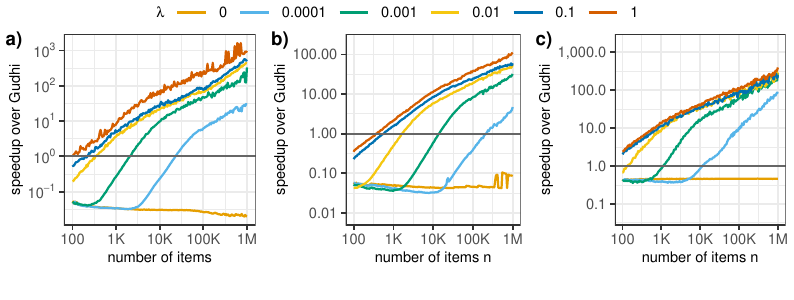}
    \caption{
        Timing the maintenance of banana trees relative to \Gudhiname{} on linear-time inputs with varying noise parameter $\lambda$.
        The baseline of no speedup ($10^0 = 1$) is shown as a \emph{horizontal gray} line.
        \refinsubcaption{fig:local-wc-speedup} Median speedup for large updates in the linear-time case for local maintenance.
        \refinsubcaption{fig:cut-wc-speedup} Median speedup in the linear-time case for cutting.
            The peaks in the bottom right for $\lambda = 0$ (yellow line) originate in the running time for \Gudhiname{} and are not observed in the time for banana trees.
        \refinsubcaption{fig:glue-wc-speedup} Median speedup in the linear-time case for gluing.
    }
    \label{fig:linear-time-speedup}
\end{figure}
\Cref{fig:linear-time-speedup} shows speedup over \Gudhiname{} in the worst cases for local and topological maintenance at varying levels of noise.

\FloatBarrier
\subsection{Quasi-Periodic Inputs}
\begin{figure}[ht]
    \centering
    \begin{subfigure}{0\linewidth}\phantomcaption\label{fig:quasi-periodic-construction-vs-n}\end{subfigure}
    \begin{subfigure}{0\linewidth}\phantomcaption\label{fig:quasi-periodic-construction-vs-sigma}\end{subfigure}
    \includegraphics{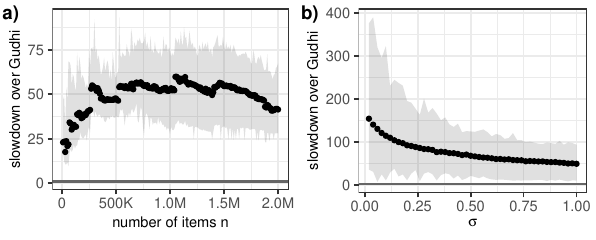}
    \caption{
        Timing the construction of banana trees for quasi-periodic inputs. Both plots show the slowdown over \Gudhiname{}.
        Ribbons extend from the minimum to the maximum observed value; dots show the mean.
        The baseline of no slowdown is shown as a \emph{horizontal gray} line.
        \refinsubcaption{fig:quasi-periodic-construction-vs-n} Slowdown for $\sigma = 1$ dependent on input size $n$.
        \refinsubcaption{fig:quasi-periodic-construction-vs-sigma} Slowdown dependent on $\sigma$ and averaged over all input sizes.
    }
    \label{fig:quasi-periodic-construction}
\end{figure}
The slowdowns of our construction algorithm over \Gudhiname{} on quasi-periodic inputs are similar to those on the random walks:
For $\sigma = 1$ we observe average values between $17.5$ and $54.1$.
Reducing the intensity of the noise by reducing $\sigma$ increases the slowdown to $154$ at $\sigma = 0.02$ on average over all input sizes.
With smaller $\sigma$ the fraction of items that is critical also gets smaller, since the bias due to the sine wave increases relative to the standard deviation of the random walk.
Reducing the number of critical items appears to improve the performance of \Gudhiname{} more so than it improves the performance of our construction algorithm,
as we also observed for the construction of random walks.
This also explains the greater slowdowns for $\sigma = 1$ compared to unbiased random walks, as the number of critical items is lower than on random walks (\qty{37.4}{\percent} compared to \qty{50}{\percent}).

\begin{figure}
    \centering
    \includegraphics{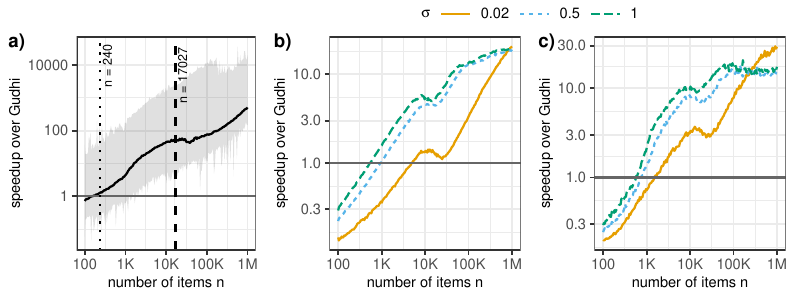}
    \begin{subfigure}{0\linewidth}\phantomcaption\label{fig:mqp-local-speedup-vs-n}\end{subfigure}
    \begin{subfigure}{0\linewidth}\phantomcaption\label{fig:mqp-cut-speedup}\end{subfigure}
    \begin{subfigure}{0\linewidth}\phantomcaption\label{fig:mqp-glue-speedup}\end{subfigure}
    \caption{
        Speedup of banana trees relative to \Gudhiname{} when performing local and topological updates on quasi-periodic inputs.
        The baseline of no speedup is shown as a \emph{horizontal gray} line.
        \refinsubcaption{fig:mqp-local-speedup-vs-n}
            Speedup when performing local updates to inputs of size $n$ and with parameter $\sigma = 1$.
            The ribbon extends from the minimum to the maximum observed value; the line traces the median speedup.
            Banana trees are faster than \Gudhiname{} on all inputs with $n \geq 1683$; see \emph{vertical dashed} line.
            In the median they are faster on all inputs with $n \geq 264$, as shown by the \emph{dotted} line.
        \refinsubcaption{fig:mqp-cut-speedup} Speedup of cutting inputs of size $n$ with parameter $\sigma$ in half.
        \refinsubcaption{fig:mqp-glue-speedup} Speedup of gluing two equally large inputs with parameter $\sigma$ into a list of size $n$.
    }
    \label{fig:mqp-local-speedup-combined}
\end{figure}
\Cref{fig:mqp-local-speedup-combined} shows the speedup of banana trees over \Gudhiname{} for local and topological maintenance.
Compare to \cref{fig:maintenance-speedup} and notice the similarities to the speedup for unbiased random walks ($\mu = 1$).

\FloatBarrier
\subsection{Real-World Examples}
The definition of persistence diagrams requires unique function values of the critical items.
However, this uniqueness is not guaranteed in real-world data.
As mentioned in \cref{sec:real-world-examples}, we achieve this by adding noise.
Note that, in the case where multiple consecutive items in a time series have identical values,
this would add new critical items.
Hence, we first replace any such consecutive items by a single item.
We then add to each item independently a random number uniformly distributed in $[-0.5\cdot 10^{-2}\cdot \Delta f, 0.5 \cdot 10^{-2} \Delta f]$, 
where $\Delta f$ is the smallest non-zero difference between any two items of $f$.

\end{document}

%% file: graphics/tikz/windows.tex
\newcommand{\drawwindow}[5]{
    \draw[thick] (#1,#2) -- (#1, #4) -- (#5,#4) -- (#5,#2)  -- cycle;
    \draw[dotted, thick] (#3,#4) -- (#3,#2);
}
\begin{tikzpicture}[scale=\windowbananascale]
\input{graphics/tikz/windows-coords}

\coordinate (o) at (\xo, \yo);
\coordinate (p) at (\xp, \yp);
\coordinate (q) at (\xq, \yq);
\coordinate (r) at (\xr, \yr);
\coordinate (s) at (\xs, \ys);
\coordinate (t) at (\xt, \yt);
\coordinate (u) at (\xu, \yu);
\coordinate (v) at (\xv, \yv);
\coordinate (w) at (\xw, \yw);
\coordinate (x) at (\xx, \yx);
\coordinate (y) at (\xy, \yy);

\draw[very thick,oiblue]
    (o) .. controls ++(0,0) and ++(-0.6,0) ..
    (p) .. controls ++(\afterxoff,0) ..
    (\xqrin, \yr) .. controls ++(0,0) and ++(\beforexoff,0) ..
    (q) .. controls ++(\afterxoff,0) and ++(\beforexoff,0) ..
    (r) .. controls ++(\afterxoff,0) ..
    (\xstin, \yt) .. controls ++(0,0) and ++(\beforexoff,0) ..
    (s) .. controls ++(\afterxoff,0) and ++(\beforexoff,0) ..
    (t) .. controls ++(\afterxoff,0) and ++(\beforexoff,0) ..
    (u) .. controls ++(\afterxoff,0) and ++(\beforexoff,0) ..
    (v) .. controls ++(\afterxoff,0) and ++(\beforexoff,0) ..
    (w) .. controls ++(\afterxoff,0) ..
    (\xwvin, \yv) --
    (x) --
    (y);

\drawwindow{\xp}{\yp}{\xu}{\yu}{\xupin}
\drawwindow{\xr}{\yr}{\xq}{\yq}{\xqrin}
\drawwindow{\xt}{\yt}{\xs}{\ys}{\xstin}
\drawwindow{\xv}{\yv}{\xw}{\yw}{\xwvin}

\begin{scope}
    \draw[->] (\xu,\yp) ++(-0.2,0.3) -- ++(-3,0) node[midway, above] {{\tiny mid-panel}};
    \draw[->] (\xu,\yp) ++( 0.2,0.3) -- ++( 3,0) node[midway, above] {{\tiny in-panel}};
\end{scope}

\begin{scope}[every circle/.style={radius=0.1}, draw=oiblue, very thick, fill=oicyan]
    \node[above] at (p) {$a$};
    \fill (p) circle; \draw (p) circle;
    \node[below] at (q) {$q$};
    \fill (q) circle; \draw (q) circle;
    \node[above] at (r) {$r$};
    \fill (r) circle; \draw (r) circle;
    \node[below] at (s) {$s$};
    \fill (s) circle; \draw (s) circle;
    \node[above] at (t) {$t$};
    \fill (t) circle; \draw (t) circle;
    \node[below] at (u) {$b$};
    \fill (u) circle; \draw (u) circle;
    \node[above] at (v) {$v$};
    \fill (v) circle; \draw (v) circle;
    \node[below] at (w) {$w$};
    \fill (w) circle; \draw (w) circle;
\end{scope}

\end{tikzpicture}

%% file: graphics/tikz/bananas.tex
\begin{tikzpicture}[scale=\windowbananascale]
\input{graphics/tikz/windows-coords}

\tikzmath{
    \xoffset1 = 0.75;
    \xoffset2 = 1;
    \xp = 0; \xu = 0;
    \xr = \xp - \xoffset1;
    \xt = \xp - \xoffset1;
    \xq = \xr - \xoffset2;
    \xs = \xt - \xoffset2;
    \xv = \xu + \xoffset1;
    \xw = \xv + \xoffset2;
}

\coordinate (p) at (\xp,\yp);
\coordinate (q) at (\xq,\yq);
\coordinate (r) at (\xr,\yr);
\coordinate (s) at (\xs,\ys);
\coordinate (t) at (\xt,\yt);
\coordinate (u) at (\xu,\yu);
\coordinate (v) at (\xv,\yv);
\coordinate (w) at (\xw,\yw);

\begin{scope}[every path/.style={thick, dashed, oiblue}]
    \draw[bend left=10] (p) to ++(195:2.2);
    \draw[bend right=10] (p) to ++( 45:2.2);
\end{scope}

\begin{scope}[every path/.style={very thick, oiblue}]
    \drawemptybanana{r}{q}
    \drawemptybanana{t}{s}
    \drawemptybanana{v}{w}
    \draw[bend right, out = 220, in = 90, relative = false] (p) to (r);
    \draw (r) -- (t);
    \draw[bend right, out = 270, in = 135, relative = false] (t) to (u);
    \draw[bend left, out = -45, in = 90, relative = false] (p) to (v);
    \draw[bend left, out = 270, in = 45, relative = false] (v) to (u);
\end{scope}

\begin{scope}[every circle/.style={radius=0.1}, draw=oiblue, very thick, fill=oicyan]
    \fill (p) circle; \draw (p) circle;
    \node[above] at (p) {$a$};
    \fill (q) circle; \draw (q) circle;
    \node[below] at (q) {$q$};
    \fill (r) circle; \draw (r) circle;
    \node[right] at (r) {$r$};
    \fill (s) circle; \draw (s) circle;
    \node[below] at (s) {$s$};
    \fill (t) circle; \draw (t) circle;
    \node[right] at (t) {$t$};
    \fill (u) circle; \draw (u) circle;
    \node[below] at (u) {$b$};
    \fill (v) circle; \draw (v) circle;
    \node[left] at (v) {$v$};
    \fill (w) circle; \draw (w) circle;
    \node[below] at (w) {$w$};
\end{scope}
    
\end{tikzpicture}

%% file: graphics/tikz/banana-pointers.tex
\begin{tikzpicture}[scale=\windowbananascale]
\input{graphics/tikz/windows-coords}
\tikzmath{
    \pointerlength = 2.5;
    \xlowa = \yp - \yo;
}

\coordinate (a) at ( 0,  \yp);
\coordinate (b) at (-4.5,  \yu);
\coordinate (lowa) at (\xlowa, \yo);

\begin{scope}[every path/.style={thick}]
    \draw[->,oiorange] (a) -- ++(205:\pointerlength) node [above left] {$\inptr$};
    \draw[->,oigreen] (a) -- ++(245:\pointerlength) node [right] {$\midptr$};
    \draw[->,oiorange] (b) -- ++( 65:\pointerlength) node [left] {$\inptr$};
    \draw[->,oigreen] (b) -- ++( 25:\pointerlength) node [below] {$\midptr$};
    \draw[->,oiblue] (a) -- ++(-55:\pointerlength) node [midway, right] {$\dnptr$};
    \draw[->,oiblue] (a) -- ++(100:\pointerlength) node [right] {$\upptr$};
\end{scope}

\begin{scope}[every path/.style={dotted, thick}]
    \path (a) ++(225:0.3) coordinate (leftbelowa);
    \draw[->] (b) -- (leftbelowa) node [midway] {$\dthptr$};

    \path (b) ++(225:0.2) coordinate (leftbelowb);

    \path (lowa) ++(90:0.2) coordinate (abovelowa);
    \draw[->, bend left=50] (a) to node [midway, right] {$\lowptr$} (abovelowa);
\end{scope}

\begin{scope}[every path/.style={dashed, thick}]
    \path (b) ++(65:\pointerlength) coordinate (inb);
    \draw[bend right=15,oiorange] (a) ++(205:\pointerlength) to (inb);
    \path (b) ++(25:\pointerlength) coordinate (midb);
    \draw[bend left=15,oigreen] (a) ++(245:\pointerlength) to (midb);

    \draw[bend right=10,oiblue] (a) ++(-55:\pointerlength) to (lowa);
\end{scope}

\begin{scope}[every circle/.style={radius=0.1}, draw = oiblue, fill = oicyan, very thick]
    \fill (a) circle; \draw (a) circle;
    \node[above right] at (a) {$a$};
    \fill (b) circle; \draw (b) circle;
    \node[below left] at (b) {$b$};
    \fill (lowa) circle; \draw (lowa) circle;
    \node[below] at (lowa) {$\lowptr(a)$};
\end{scope}

\end{tikzpicture}

%% file: graphics/tikz/anticancellation-function.tex
\begin{tikzpicture}[scale=\actikzscale]
\input{graphics/tikz/anticancellation-coords}

\coordinate (n) at (\xn,\yn);
\coordinate (o) at (\xo,\yo);
\coordinate (p) at (\xp,\yp);
\coordinate (q) at (\xq,\yq);
\coordinate (qpost) at (\xq,\yqpost);
\coordinate (r) at (\xr,\yr);
\coordinate (s) at (\xs,\ys);
\coordinate (t) at (\xt,\yt);
\coordinate (u) at (\xu,\yu);
\coordinate (v) at (\xv,\yv);
\coordinate (w) at (\xw,\yw);
\coordinate (x) at (\xx,\yx);
\coordinate (y) at (\xy,\yy);

\draw[very thick, oiblue]
    (n) .. controls ++(0,0) and ++(-\splineoffset,0) ..
    (o) .. controls ++(\splineoffset,0) and ++(0,0) ..
    (p) --
    (q) .. controls ++(0,0) and ++(-\splineoffset,0) ..
    (r) .. controls ++(\splineoffset,0) and ++(-\splineoffset,0) ..
    (s) .. controls ++(\splineoffset,0) and ++(-\splineoffset,0) ..
    (t) .. controls ++(\splineoffset,0) and ++(-\splineoffset,0) ..
    (u) .. controls ++(\splineoffset,0) and ++(-\splineoffset,0) ..
    (v) .. controls ++(\splineoffset,0) and ++(-\splineoffset,0) ..
    (w) .. controls ++(\splineoffset,0) and ++(-\splineoffset,0) ..
    (x) .. controls ++(\splineoffset,0) ..
    (y);

\draw[very thick, oired, dotted]
    (p) .. controls ++(\splineoffset, 0) and ++(-\splineoffset,0) ..
    (qpost) .. controls ++(\splineoffset, 0) and ++(-\splineoffset,0) ..
    (r);
\draw[<-, oired, thick] (qpost) ++(0,-0.2) -- (q);

\begin{scope}[every circle/.style={radius=0.1}, draw = oiblue, fill = oicyan, thick]
    \fill (o) circle; \draw (o) circle;
    \fill (p) circle; \draw (p) circle;
    \fill (q) circle; \draw (q) circle;
    \fill[oiorange] (qpost) circle; \draw[oired] (qpost) circle;
    \fill (r) circle; \draw (r) circle;
    \fill (s) circle; \draw (s) circle;
    \fill (t) circle; \draw (t) circle;
    \fill (u) circle; \draw (u) circle;
    \fill (v) circle; \draw (v) circle;
    \fill (w) circle; \draw (w) circle;
    \fill (x) circle; \draw (x) circle;

    \node[above] at (o) {$m$};
    \node[left] at (p)  {$k$};
    \node[left] at (q)  {$j$};
    \node[below] at (r) {$r$};
    \node[above] at (s) {$s$};
    \node[below] at (t) {$t$};
    \node[above] at (u) {$u$};
    \node[below] at (v) {$v$};
    \node[above] at (w) {$w$};
    \node[below] at (x) {$x$};
\end{scope}
    
\end{tikzpicture}

%% file: graphics/tikz/anticancellation-tree.tex
\begin{tikzpicture}[scale=\actikzscale]
\input{graphics/tikz/anticancellation-coords}

\coordinate (o) at (-0.5,\yo);
\coordinate (p) at (-5.0,\yp);
\coordinate (q) at (-3.25,\yqpost);
\coordinate (r) at (-6,\yr);
\coordinate (s) at (-4.5,\ys);
\coordinate (t) at (-4,\yt);
\coordinate (u) at (-2.5,\yu);
\coordinate (v) at (-2.0,\yv);
\coordinate (w) at (-1.5,\yw);
\coordinate (x) at (0.5,\yx);

\begin{scope}[every path/.style={very thick, oiblue}]
    \draw[bend right=10] (o) to (w);
    \draw[bend right=10] (w) to (x);
    \draw[bend left=10] (o) to (x);

    \draw[bend right=10] (w) to (u);
    \draw[bend right=10] (u) to (v);
    \draw[bend left=10] (w) to (v);

    \draw[bend right=10] (u) to (q) to (s) to (t);
    \draw[bend left=10] (u) to (t);

    \drawemptybanana{s}{r};

    \draw[bend left, oired] (p) to (q);
    \draw[bend right, oired] (p) to (q);
\end{scope}

\begin{scope}[every path/.style = {oired, dotted, very thick}]
    \path (w) ++(135:0.1) coordinate (aboveleftw);
    \path (u) ++(135:0.1) coordinate (aboveleftu);
    \path (q) ++(120:0.1) coordinate (aboveleftq);

    \draw[->,bend right = 90, looseness = 2] (o) to (aboveleftw);
    \draw[->,bend right = 90, looseness = 2] (w) to (aboveleftu);
    \draw[->,out=-90, in=250, relative, looseness = 2] (u) to (aboveleftq);
\end{scope}

\begin{scope}[every circle/.style={radius=0.1}, draw = oiblue, fill = oicyan, thick]
    \fill (o) circle; \draw (o) circle;
    \fill[oiorange] (p) circle; \draw[oired] (p) circle;
    \fill[oiorange] (q) circle; \draw[oired] (q) circle;
    \fill (r) circle; \draw (r) circle;
    \fill (s) circle; \draw (s) circle;
    \fill (t) circle; \draw (t) circle;
    \fill (u) circle; \draw (u) circle;
    \fill (v) circle; \draw (v) circle;
    \fill (w) circle; \draw (w) circle;
    \fill (x) circle; \draw (x) circle;

    \node[above] at (o) {$m$};
    \node[left] at (p) {$k$};
    \node[below] at (q) {$j$};
    \node[below] at (r) {$r$};
    \node[above left] at (s) {$s$};
    \node[below] at (t) {$t$};
    \node[below right] at (u) {$u$};
    \node[below] at (v) {$v$};
    \node[right] at (w) {$w$};
    \node[below] at (x) {$x$};
\end{scope}

\end{tikzpicture}

%% file: graphics/tikz/max-interchange-1.tex
\begin{tikzpicture}[scale=\maxxchangetikzscale]

\tikzmath{
    \scopeshift = 5;
    \rightscopeleft = \scopeshift - 1.00;
}

\coordinate (p) at (-0.5, 2);
\coordinate (q) at ( 1.0, 3);
\coordinate (i) at ( 0.0, 0.0);
\coordinate (ipost) at ( 1.0, 0.0);
\coordinate (jpre) at ( 1.5, 1);
\coordinate (jpost) at ( 1.5, 3.5);

\begin{scope}[every path/.style={very thick}]
    \draw[bend right, oired] (q) to (p);
    \draw[bend left, oigreen] (q) to (p);
    
    \draw[bend right, oired, dotted] (jpre) to (i);
    \draw[bend left, oigreen, dotted] (jpre) to (i);

    \draw[bend left=10, oiblue, dashed] (q) to ++(100:1.5);
    \draw[bend right=10, oiblue] (q) to (jpre);
    \draw[bend right=10, oiblue, dashed] (jpre) to ++(300:1.5);
\end{scope}

\begin{scope}[every circle/.style={radius=0.1}, fill=oicyan, draw=oiblue, very thick]
    \fill (p) circle; \draw (p) circle;
    \fill (q) circle; \draw (q) circle;
    \fill (i) circle; \draw (i) circle;
    \fill (jpre) circle; \draw (jpre) circle;
    \node[left] at (p) {$p$};
    \node[right] at (q) {$q$};
    \node[left] at (i) {$i$};
    \node[right] at (jpre) {$j$};
\end{scope}

\draw[->, thick] (2.5, 2.0) -- (\rightscopeleft, 2.0);

\tikzset{*/.style={shift={(\scopeshift,0)}}}
\begin{scope}[every path/.style={very thick}, shift = {(\scopeshift,0)}]
    \draw[bend right, oired] ([*]q) to ([*]p);
    \draw[bend left, oigreen] ([*]q) to ([*]p);
    
    \draw[bend right=10, oired, dotted] ([*]q) to ([*]ipost);
    \draw[bend left=10, oigreen, dotted] ([*]jpost) to ([*]ipost);

    \draw[bend left=10, oiblue, dashed] ([*]jpost) to ++(100:1.0);
    \draw[bend left=30, oiblue] ([*]q) to ([*]jpost);
    \draw[bend right=10, oiblue, dashed] ([*]jpost) to ++(300:1.5);
\end{scope}

\begin{scope}[every circle/.style={radius=0.1}, fill=oicyan, draw=oiblue, very thick, shift = {(\scopeshift,0)}]
    \fill ([*]p) circle; \draw ([*]p) circle;
    \fill ([*]q) circle; \draw ([*]q) circle;
    \fill ([*]ipost) circle; \draw ([*]ipost) circle;
    \fill ([*]jpost) circle; \draw ([*]jpost) circle;
    \node[below] at ([*]p) {$p$};
    \node[above left] at ([*]q) {$q$};
    \node[left] at ([*]ipost) {$i$};
    \node[right] at ([*]jpost) {$j$};
\end{scope}

\end{tikzpicture}

%% file: graphics/tikz/max-interchange-2.tex
\begin{tikzpicture}[scale=\maxxchangetikzscale]

\tikzmath{
    \scopeshift = 5;
    \rightscopeleft = \scopeshift - 1.00;
}

\coordinate (p) at (-1, 0);
\coordinate (q) at ( 1.0, 2.5);
\coordinate (i) at ( 0.5, 0.5);
\coordinate (ipost) at ( 1.5, 0.5);
\coordinate (jpre) at ( 1.5, 1);
\coordinate (jpost) at ( 1.5, 3.5);

\begin{scope}[every path/.style={very thick}]
    \draw[bend right, oired] (q) to (p);
    \draw[bend left=15, oigreen] (q) to (p);
    
    \draw[bend right, oired, dotted] (jpre) to (i);
    \draw[bend left, oigreen, dotted] (jpre) to (i);

    \draw[bend left=10, oiblue, dashed] (q) to ++(100:1.5);
    \draw[bend right=10, oiblue] (q) to (jpre);
    \draw[bend right=10, oiblue, dashed] (jpre) to ++(300:1.5);
\end{scope}

\begin{scope}[every circle/.style={radius=0.1}, fill=oicyan, draw=oiblue, very thick]
    \fill (p) circle; \draw (p) circle;
    \fill (q) circle; \draw (q) circle;
    \fill (i) circle; \draw (i) circle;
    \fill (jpre) circle; \draw (jpre) circle;
    \node[left] at (p) {$p$};
    \node[right] at (q) {$q$};
    \node[below] at (i) {$i$};
    \node[right] at (jpre) {$j$};
\end{scope}

\draw[->, thick] (2.5, 2.0) -- (\rightscopeleft, 2.0);

\tikzset{*/.style={shift={(\scopeshift,0)}}}
\begin{scope}[every path/.style={very thick}, shift = {(\scopeshift,0)}]
    \draw[bend right, oired] ([*]jpost) to ([*]p);
    \draw[bend left=15, oigreen] ([*]q) to ([*]p);
    
    \draw[bend right=20, oired, dotted] ([*]q) to ([*]ipost);
    \draw[bend left=20, oigreen, dotted] ([*]q) to ([*]ipost);

    \draw[bend left=10, oiblue, dashed] ([*]jpost) to ++(100:1.0);
    \draw[bend right=10, oiblue] ([*]q) to ([*]jpost);
    \draw[bend right=10, oiblue, dashed] ([*]jpost) to ++(300:1.5);
\end{scope}

\begin{scope}[every circle/.style={radius=0.1}, fill=oicyan, draw=oiblue, very thick, shift = {(\scopeshift,0)}]
    \fill ([*]p) circle; \draw ([*]p) circle;
    \fill ([*]q) circle; \draw ([*]q) circle;
    \fill ([*]ipost) circle; \draw ([*]ipost) circle;
    \fill ([*]jpost) circle; \draw ([*]jpost) circle;
    \node[left] at ([*]p) {$p$};
    \node[left] at ([*]q) {$q$};
    \node[below] at ([*]ipost) {$i$};
    \node[right] at ([*]jpost) {$j$};
\end{scope}

\end{tikzpicture}

%% file: graphics/tikz/min-interchange.tex
\begin{tikzpicture}[scale=\maxxchangetikzscale]

\tikzmath{
    \scopeshift = 4;
    \rightscopeleft = \scopeshift - 1.00;
}

\coordinate (ppre) at  ( 0.0, 0.0);
\coordinate (ppost) at ( 1.5, 0.25);
\coordinate (q) at     ( 0.0, 3.5);
\coordinate (ipre) at  (-1.5, 0.25);
\coordinate (ipost) at ( 0.0, 0.0);
\coordinate (jpre) at  (-0.5, 1.75);
\coordinate (jpost) at ( 0.5, 1.75);

\coordinate (b) at (0.5, 2.50);
\coordinate (apre) at (0.5, 1.00);
\coordinate (apost) at (1.25, 1.25);

\begin{scope}[every path/.style={very thick}]
    \draw[bend right=15, oired] (q) to (jpre);
    \draw[bend right=15, oired, dotted] (jpre) to (ppre);
    \draw[bend left=15, oigreen] (q) to (b);
    \draw[bend left=15, oiblue] (b) to (apre);
    \draw[bend left=15, oigreen, dotted] (apre) to (ppre);
    
    \draw[bend right, oired] (jpre) to (ipre);
    \draw[bend left, oigreen] (jpre) to (ipre);
\end{scope}

\begin{scope}[every circle/.style={radius=0.1}, fill=oicyan, draw=oiblue, very thick]
    \fill (ppre) circle; \draw (ppre) circle;
    \fill (q) circle; \draw (q) circle;
    \fill (ipre) circle; \draw (ipre) circle;
    \fill (jpre) circle; \draw (jpre) circle;
    \fill (b) circle; \draw (b) circle;
    \fill (apre) circle; \draw (apre) circle;
    \node[left] at (ppre) {$p$};
    \node[above] at (q) {$q$};
    \node[left] at (ipre) {$i$};
    \node[right] at (jpre) {$j$};
    \node[right] at (b) {$b$};
    \node[right] at (apre) {$a$};
\end{scope}

\draw[->, thick] (1.5, 2.0) -- (\rightscopeleft, 2.0);

\tikzset{*/.style={shift={(\scopeshift,0)}}}
\begin{scope}[every path/.style={very thick}, shift = {(\scopeshift,0)}]
    \draw[bend right, oired] ([*]q) to ([*]ipost);
    \draw[bend left=15, oigreen] ([*]q) to ([*]b);
    \draw[bend left=15, oiblue] ([*]b) to ([*]jpost);
    \draw[bend left=15, oigreen] ([*]jpost) to ([*]ipost);
    
    \draw[bend right, oired, dotted] ([*]jpost) to ([*]ppost);
    \draw[bend left=10, oiblue] ([*]jpost) to ([*]apost);
    \draw[bend left=10, oigreen, dotted] ([*]apost) to ([*]ppost);
\end{scope}

\begin{scope}[every circle/.style={radius=0.1}, fill=oicyan, draw=oiblue, very thick, shift = {(\scopeshift,0)}]
    \fill ([*]ppost) circle; \draw ([*]ppost) circle;
    \fill ([*]q) circle; \draw ([*]q) circle;
    \fill ([*]ipost) circle; \draw ([*]ipost) circle;
    \fill ([*]jpost) circle; \draw ([*]jpost) circle;
    \fill ([*]b) circle; \draw ([*]b) circle;
    \fill ([*]apost) circle; \draw ([*]apost) circle;
    \node[right] at ([*]ppost) {$p$};
    \node[above] at ([*]q) {$q$};
    \node[left] at ([*]ipost) {$i$};
    \node[left] at ([*]jpost) {$j$};
    \node[right] at ([*]b) {$b$};
    \node[above right] at ([*]apost) {$a$};
\end{scope}

\end{tikzpicture}

%% file: graphics/tikz/injury.tex
\begin{tikzpicture}[scale=\topotikzscale]
\tikzmath{ \scopeshift = 7; }

\coordinate (a) at ( 0,  0.0);
\coordinate (b) at (-1,  2.0);
\coordinate (j) at ( 1.5, -0.5);
\coordinate (k) at ( 2.5,  1.5);

\path[bend left] (1.5, -0.5) to coordinate[midway] (cut) (2.5, 1.5) ;

\begin{scope}[draw=oiblue, very thick]
    \drawemptybanana{a}{b}

    \path (cut) -- ++(240:0.05) coordinate (leftbelowcut);
    \path (cut) -- ++(60:0.05) coordinate (rightabovecut);
    \draw[oired, bend left=15] (j) to (leftbelowcut);
    \draw[oigreen, bend left=15, dashed] (rightabovecut) to coordinate[midway] (betweenjk) (k);
    \draw[oired, bend right] (j) to (k);

    \draw[black, thick] (leftbelowcut) ++(150:0.1) -- ++(-30:0.2);
    \draw[black, thick] (rightabovecut) ++(150:0.1) -- ++(-30:0.2);

    \draw[bend left=10, dashed] (b) to ++(220:1);
    \draw[bend right=10, dashed] (b) to ++(70:1);

    \draw[bend left=10, dashed] (k) to ++(120:1);
    \draw[bend right=10, dashed] (k) to ++(-30:1);
\end{scope}

\begin{scope}
    \path (a) -- ++(120:0.3) coordinate (endofarrow);
    \draw[dotted, oigreen, very thick, bend right, ->] (betweenjk) ++(120:0.1) to (endofarrow);
\end{scope}

\begin{scope}[every circle/.style={radius=0.1}, draw=oiblue, very thick, fill=oicyan]
    \fill (a) circle; \draw (a) circle;
    \node[below] at (a) {$a$};
    \fill (b) circle; \draw (b) circle;
    \node[above left] at (b) {$b$};
    \fill (j) circle; \draw (j) circle;
    \node[left] at (j) {$j$};
    \fill (k) circle; \draw (k) circle;
    \node[above right] at (k) {$k$};
\end{scope}

\draw[->, thick] (3.75, 0.75) -- (4.75, 0.75);

\tikzset{*/.style={shift={(\scopeshift,0)}}}
\begin{scope}[draw=oiblue, very thick, shift={(\scopeshift,0)}]
    \draw[bend right] ([*]a) to ([*]b); 
    \path[bend left] ([*]a) to coordinate[midway] (aftercut) ([*]b);
    \draw[bend left=15, oigreen, dashed] ([*]a) to (aftercut);
    \draw[bend left=15] (aftercut) to ([*]b);

    \path ([*]cut) -- ++(240:0.05) coordinate ([*]leftbelowcut);
    \path ([*]cut) -- ++(60:0.05) coordinate ([*]rightabovecut);

    \draw[bend left, oired] ([*]j) to ([*]k);
    \draw[bend right, oired] ([*]j) to ([*]k);

    \draw[bend left=10, dashed] ([*]b) to ++(220:1);
    \draw[bend right=10, dashed] ([*]b) to ++(70:1);

    \draw[bend left=10, dashed] ([*]k) to ++(120:1);
    \draw[bend right=10, dashed] ([*]k) to ++(-30:1);
\end{scope}

\begin{scope}[every circle/.style={radius=0.1}, draw=oiblue, very thick, fill=oicyan, shift={(\scopeshift,0)}]
    \fill ([*]a) circle; \draw ([*]a) circle;
    \node[below] at ([*]a) {$a$};
    \fill ([*]b) circle; \draw ([*]b) circle;
    \node[above left] at ([*]b) {$b$};
    \fill ([*]j) circle; \draw ([*]j) circle;
    \node[left] at ([*]j) {$j$};
    \fill ([*]k) circle; \draw ([*]k) circle;
    \node[above right] at ([*]k) {$k$};
\end{scope}

\end{tikzpicture}

%% file: graphics/tikz/fatality.tex
\begin{tikzpicture}[scale=\topotikzscale]
\tikzmath{ \scopeshift = 7; }

\coordinate (a) at ( 0,  0.0);
\coordinate (b) at (-1,  2.0);
\coordinate (j) at ( 1.5, -0.5);
\coordinate (k) at ( 2.5,  1.5);

\path[bend right] (1.5, -0.5) to coordinate[midway] (cut) (2.5, 1.5) ;
\path[bend left] (1.5, -0.5) to coordinate[midway] (arrowstart) (2.5, 1.5) ;

\begin{scope}[draw=oiblue, very thick]
    \drawemptybanana{a}{b}

    \path (cut) -- ++(240:0.05) coordinate (leftbelowcut);
    \path (cut) -- ++(60:0.05) coordinate (rightabovecut);
    \draw[oigreen, bend right=15, dashed] (j) to (leftbelowcut);
    \draw[oired, bend right=15] (rightabovecut) to coordinate[midway] (betweenjk) (k);
    \draw[oigreen, bend left, dashed] (j) to (k);

    \draw[black, thick] (leftbelowcut) ++(150:0.1) -- ++(-30:0.2);
    \draw[black, thick] (rightabovecut) ++(150:0.1) -- ++(-30:0.2);

    \draw[bend left=10, dashed] (b) to ++(220:1);
    \draw[bend right=10, dashed] (b) to ++(70:1);

    \draw[bend left=10, dashed] (k) to ++(120:1);
    \draw[bend right=10, dashed] (k) to ++(-30:1);
\end{scope}

\begin{scope}
    \path (a) -- ++(60:0.2) coordinate (endofarrow);
    \draw[dotted, oigreen, very thick, bend right, <->] (arrowstart) ++(120:0.1) to (endofarrow);
\end{scope}

\begin{scope}[every circle/.style={radius=0.1}, draw=oiblue, very thick, fill=oicyan]
    \fill (a) circle; \draw (a) circle;
    \node[below] at (a) {$a$};
    \fill (b) circle; \draw (b) circle;
    \node[above left] at (b) {$b$};
    \fill[oigreen!25] (j) circle; \draw[oigreen] (j) circle;
    \node[left] at (j) {$j$};
    \fill (k) circle; \draw (k) circle;
    \node[above right] at (k) {$k$};
\end{scope}

\draw[->, thick] (3.75, 0.75) -- (4.75, 0.75);

\tikzset{*/.style={shift={(\scopeshift,0)}}}
\begin{scope}[draw=oiblue, very thick, shift={(\scopeshift,0)}]
    \draw[bend right, oigreen, dashed] ([*]a) to ([*]b); 
    \path[bend left] ([*]a) to coordinate[midway] (aftercut) ([*]b);
    \draw[bend left=15, oigreen, dashed] ([*]a) to (aftercut);
    \draw[bend left=15] (aftercut) to ([*]b);

    \path ([*]cut) -- ++(240:0.05) coordinate ([*]leftbelowcut);
    \path ([*]cut) -- ++(60:0.05) coordinate ([*]rightabovecut);

    \draw[bend left, dotted] ([*]j) to ([*]k);
    \draw[bend right, oired] ([*]j) to ([*]k);

    \draw[bend left=10, dashed] ([*]b) to ++(220:1);
    \draw[bend right=10, dashed] ([*]b) to ++(70:1);

    \draw[bend left=10, dashed] ([*]k) to ++(120:1);
    \draw[bend right=10, dashed] ([*]k) to ++(-30:1);
\end{scope}

\begin{scope}[every circle/.style={radius=0.1}, draw=oiblue, very thick, fill=oicyan, shift={(\scopeshift,0)}]
    \fill[oigreen!25] ([*]a) circle; \draw[oigreen] ([*]a) circle;
    \node[below] at ([*]a) {$j$};
    \fill ([*]b) circle; \draw ([*]b) circle;
    \node[above left] at ([*]b) {$b$};
    \fill ([*]j) circle; \draw ([*]j) circle;
    \node[left] at ([*]j) {$a$};
    \fill ([*]k) circle; \draw ([*]k) circle;
    \node[above right] at ([*]k) {$k$};
\end{scope}

\end{tikzpicture}